\documentclass[preprintnumbers,10pt,nofootinbib]{revtex4}

\usepackage{amsmath,latexsym,amssymb,amsfonts}
\usepackage{graphicx,color}
\usepackage{bm}
\usepackage{microtype}
\usepackage{amsmath}
\usepackage{amssymb}
\usepackage{subfigure}
\usepackage{epstopdf}

\begin{document}


\title{\textbf{Dynamics of Bronnikov-Ellis wormhole with double-null simulation
}}

\author{
Anyuan Xu$^{a}$\footnote{{\tt 1103697451@qq.com}},
Xiao Yan Chew$^{a}$\footnote{{\tt xiao.yan.chew@just.edu.cn}}
and
Dong-han Yeom$^{b,c,d,e}$\footnote{{\tt innocent.yeom@gmail.com}}
}

\affiliation{
$^{a}$School of Science, Jiangsu University of Science and Technology, Zhenjiang 212100, China \\
$^{b}$Department of Physics Education, Pusan National University, Busan 46241, Republic of Korea\\
$^{c}$Research Center for Dielectric and Advanced Matter Physics, Pusan National University, Busan 46241, Republic of Korea\\
$^{d}$Leung Center for Cosmology and Particle Astrophysics, National Taiwan University, Taipei 10617, Taiwan \\
$^{e}$Department of Physics and Astronomy, University of Waterloo, Waterloo, ON N2L 3G1, Canada 
}

\begin{abstract}
We investigate the dynamical collapse of Bronnikov-Ellis (BE) wormhole using the double-null formalism, where its throat is characterized by the coincidence of two curves $r_{,u} = 0$ and $r_{,v} = 0$. The emission of two ingoing pulses: normal scalar and phantom fields in the wormhole spacetime reveals two distinct instability scenarios: a normal scalar field triggers gravitational collapse into a black hole where the singularity $r=0$ hidden by the event horizon ($r_{,u}=0$ and $r_{,v}=0$); while a phantom field drives inflationary expansion of wormhole, decoupling two asymptotic regions with the cosmological horizon ($r_{,u}=0$ and $r_{,v}=0$). The process of two scenarios can be accelerated by increasing the amplitude of pulses but can be delayed by increasing the wormhole's mass. Additionally, the collisions of two identical pulses from ingoing and outgoing null directions in the massless BE wormhole fail to cure the instabilities because the two scenarios can still occur, but the formation of a black hole can be delayed for the collision of normal and phantom fields. Interestingly, the strategic tuning of emission timing for outgoing phantom field to collide with ingoing normal scalar field can temporarily stabilize the wormhole throat by restoring the coincidence of $r_{,u}$ and $r_{,v}$ again after their separation. This offers us valuable insights into extending the lifetime of a traversable wormhole.
\end{abstract}

\maketitle

\newpage

\tableofcontents

\section{Introduction}

In General Relativity (GR), traversable wormholes are intriguing astrophysical objects that can connect two distant universes through a narrow tunnel with minimal surface area, known as the throat. They do not contain a singularity or an event horizon in spacetime. However, their existence typically requires violating energy conditions, often achieved by introducing exotic matter at the throat to prevent collapse. 

One of the simplest forms of exotic matter can be represented by a phantom field, which is actually a scalar field whose kinetic term has the wrong sign in the Lagrangian, where it has been used to describe the accelerated expansion of the universe \cite{Caldwell:1999ew,Carroll:2003st,Gibbons:2003yj,Hannestad:2005fg}. It also has been employed to construct the analytical solution for the prominent example of the static and spherically symmetric traversable wormhole in GR—the Bronnikov-Ellis (BE) wormhole \cite{Ellis:1973yv,Ellis:1979bh,Bronnikov:1973fh}, subsequently has been generalized to higher dimensions \cite{Torii:2013xba}, and include rotation in four dimensions \cite{Kashargin:2007mm,Kashargin:2008pk,Kleihaus:2014dla,Chew:2016epf,Chew:2018vjp} and five dimensions \cite{Dzhunushaliev:2013jja}. The generalization of BE wormholes can be easily achieved by minimally coupling a phantom field with gravitational theories, for example, a BE wormhole immersed in bosonic matter \cite{Dzhunushaliev:2014bya,Hoffmann:2018oml,Ding:2023syj}, mixed neutron star-wormhole systems \cite{Dzhunushaliev:2011xx,Aringazin:2014rva,Dzhunushaliev:2022elv} and other generalizations can be found in Refs.~\cite{Chew:2019lsa,Chew:2020svi,Chew:2021vxh,Martinez:2020hjm,DuttaRoy:2019hij,Sharma:2022dbx,Huang:2020qmn,Blazquez-Salcedo:2020nsa,Karakasis:2021tqx,Gonzalez:2022ote,Lin:2022azn,Filho:2023yly,Hao:2024hba,Gao:2024lrb,Yang:2024prm,Ding:2024qrf,Liu:2025foo}. Besides, recent construction of traversable wormholes in some gravity theories, for instance, in the Einstein-3-form \cite{Bouhmadi-Lopez:2021zwt,Barros:2018lca,Tangphati:2023uxt}, Einstein-Dirac \cite{Blazquez-Salcedo:2019uqq,Churilova:2021tgn,Wang:2022aze} and Einstein-scalar-Gauss-Bonnet \cite{Kanti:2011yv,Ibadov:2020btp,Canate:2023mge,Ilyas:2023rde} have demonstrated that the inclusion of the phantom field is no longer necessary, this has offered us the new insights on the construction of traversable wormholes in some classical gravitational theories.

However, the static BE wormhole has been shown to possess notorious unstable radial modes in four and higher dimensions when spherical perturbation is performed in the background of the wormhole spacetime and phantom field \cite{Gonzalez:2008wd,Gonzalez:2008xk,Bronnikov:2012ch,Blazquez-Salcedo:2018ipc,Torii:2013xba}. Nevertheless, the inclusion of rotation at the wormhole throat might stabilize these static wormholes, as demonstrated by some investigation of mode analysis for rotating BE wormholes in four dimensions \cite{Azad:2023iju,Azad:2024axu} and with equal angular momenta in five dimensions \cite{Dzhunushaliev:2013jja}. These studies reveal that at the onset of rotation, the unstable radial modes bifurcated from the static limit and then increased to approach zero at a critical value of the angular momentum, leaving sufficiently rapidly rotating wormholes without unstable radial modes. Besides, the whole spectra of quasinormal modes during the ringdown stage in the emission of gravitational waves for the BE wormhole has been studied \cite{Kim:2008zzj,Blazquez-Salcedo:2018ipc,Azad:2022qqn,Khoo:2024yeh,Batic:2025hgp}. 

Thus, one might wonder whether nonlinear dynamics can stabilize the unstable static wormholes. While achieving full stability remains theoretically challenging, exploring nonlinear dynamics might give us some insights into the possible suppression of instabilities, potentially extending the lifetime of wormholes. To investigate this, we employ the double-null formalism \cite{Nakonieczna:2018tih}—a numerical framework extensively used in modeling black hole formation \cite{Waugh:1986jh,Hong:2008mw,Hwang:2012nn,Chew:2023upu}, bubble dynamics \cite{Hansen:2009kn,Hwang:2012pj}, and wormhole collapse \cite{Doroshkevich:2008xm}. We study the collapse of the BE wormhole by emitting ingoing pulses of normal scalar and phantom fields to the throat. Moreover, we introduce colliding pulses—an outgoing pulse of phantom field intersecting the ingoing ones—to explore the mechanism of instability suppression. Our results reveal novel pathways for modulating wormhole stability through nonlinear pulse interactions.

The content of our paper is organized as follows: In Sec.~\ref{sec:mod}, we briefly introduce the BE wormhole in the double-null coordinates and the implementation of double-null formalism, including the derivation of partial differential equations (PDEs) and assignment of boundary conditions. We present and discuss our numerical results in Sec.~\ref{sec:res}. In Sec.~\ref{sec:con}, we conclude our work and discuss some possible outlooks from this work.

\section{\label{sec:mod}BE wormhole in the double-null formalism}

In this section, we briefly introduce the analytical solution of the BE wormhole in the double-null coordinates and then its implementation in the double-null simulation, including the assignment of boundary conditions and free parameters.

\subsection{Analytical form of the BE wormhole}

In general, we consider a massless phantom field $\phi$ to be minimally coupled with a massless scalar field $\psi$ with the proper sign of kinetic term in the Einstein-Klein-Gordon theory:
\begin{eqnarray}\label{eqn:the-lag}
S = \int dx^{4} \sqrt{-g} \left[ \frac{1}{16\pi} R + \frac{1}{2} g^{\mu\nu} \phi_{;\mu} \phi_{;\nu} - \frac{1}{2} g^{\mu\nu} \psi_{;\mu} \psi_{;\nu} \right] \,,
\end{eqnarray}
where $R$ is the Ricci scalar, and the semicolon $(;)$ denotes the covariant derivative of a function. The variation of the above action concerning the metric $g_{\mu \nu}$, $\phi$, and $\psi$ yields the Einstein equation and two Klein-Gordon equations, respectively:
\begin{eqnarray}
G_{\mu\nu} &=& 8 \pi T_{\mu\nu}, \label{eom1}  \\
\phi_{;\mu\nu}g^{\mu\nu} &=& 0,\\
\psi_{;\mu\nu}g^{\mu\nu} &=& 0, \label{eom3}
\end{eqnarray}
where the energy-momentum tensor $T_{\mu \nu}$ is given by
\begin{eqnarray}
T_{\mu\nu} = - \phi_{;\mu}\phi_{;\nu} + \frac{1}{2} \phi_{;\rho} \phi_{;\sigma} g^{\rho\sigma} g_{\mu \nu} + \psi_{;\mu}\psi_{;\nu} - \frac{1}{2} \psi_{;\rho} \psi_{;\sigma} g^{\rho\sigma} g_{\mu \nu}.
\end{eqnarray}

\begin{figure}
    \centering
    \includegraphics[trim=30mm 20mm 30mm 20mm, clip, scale=0.38]{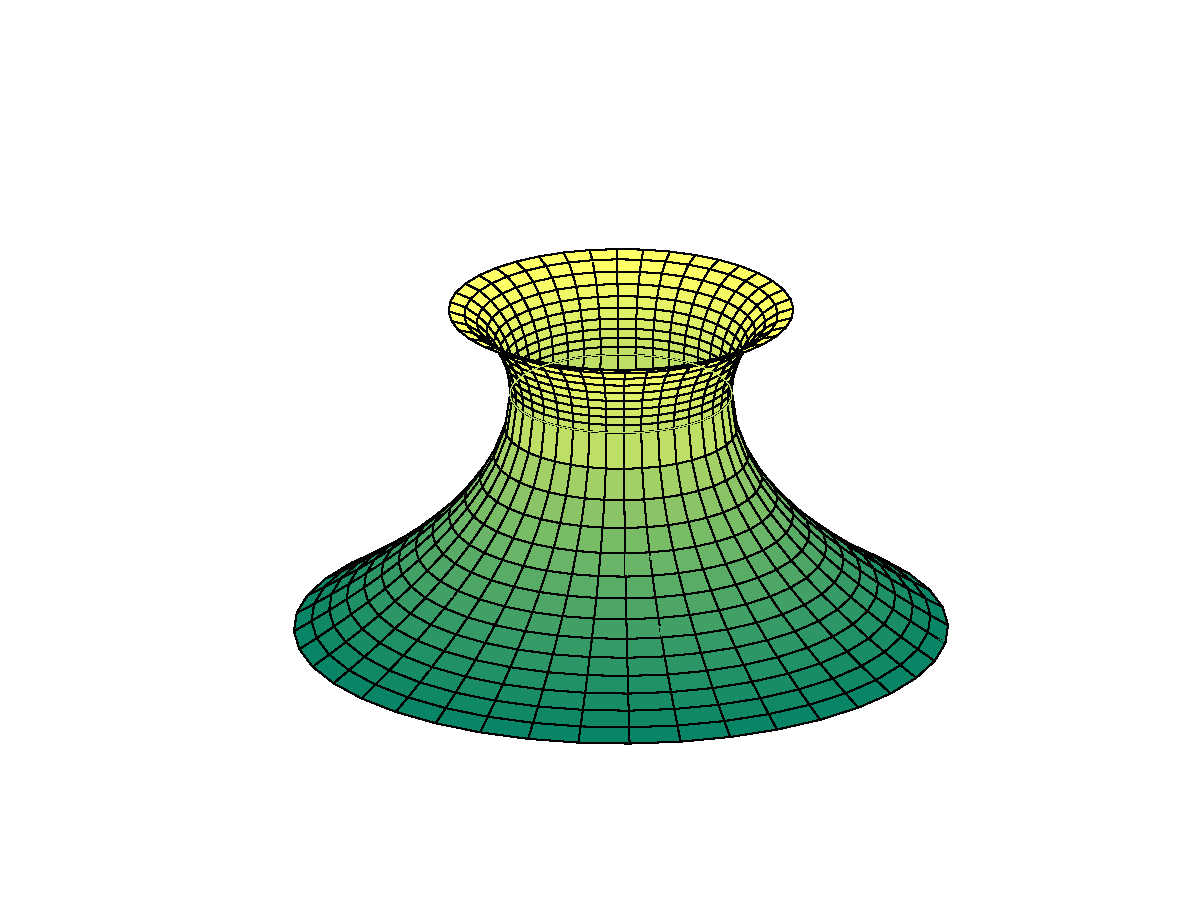}
    \includegraphics[trim=30mm 20mm 30mm 20mm, clip, scale=0.38]{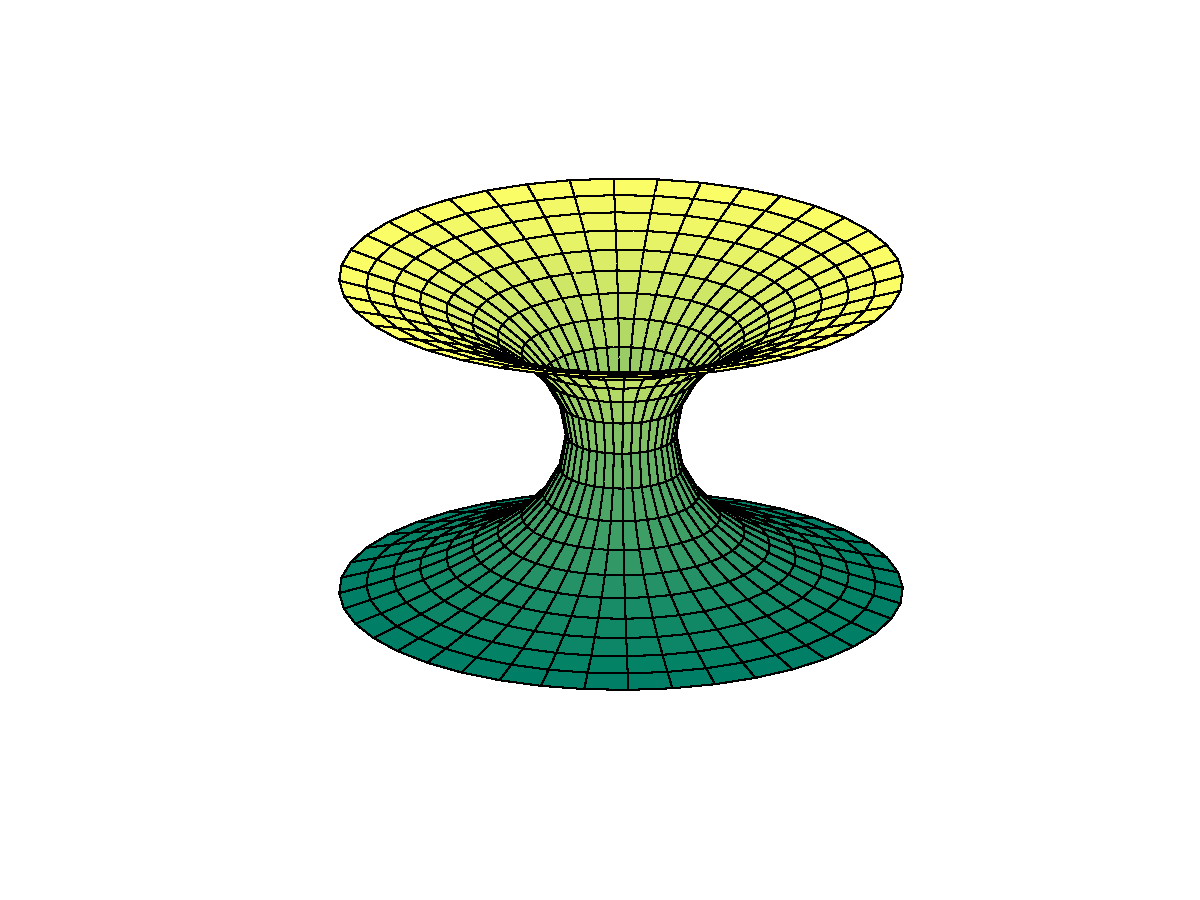}
    \includegraphics[trim=30mm 20mm 30mm 20mm, clip, scale=0.38]{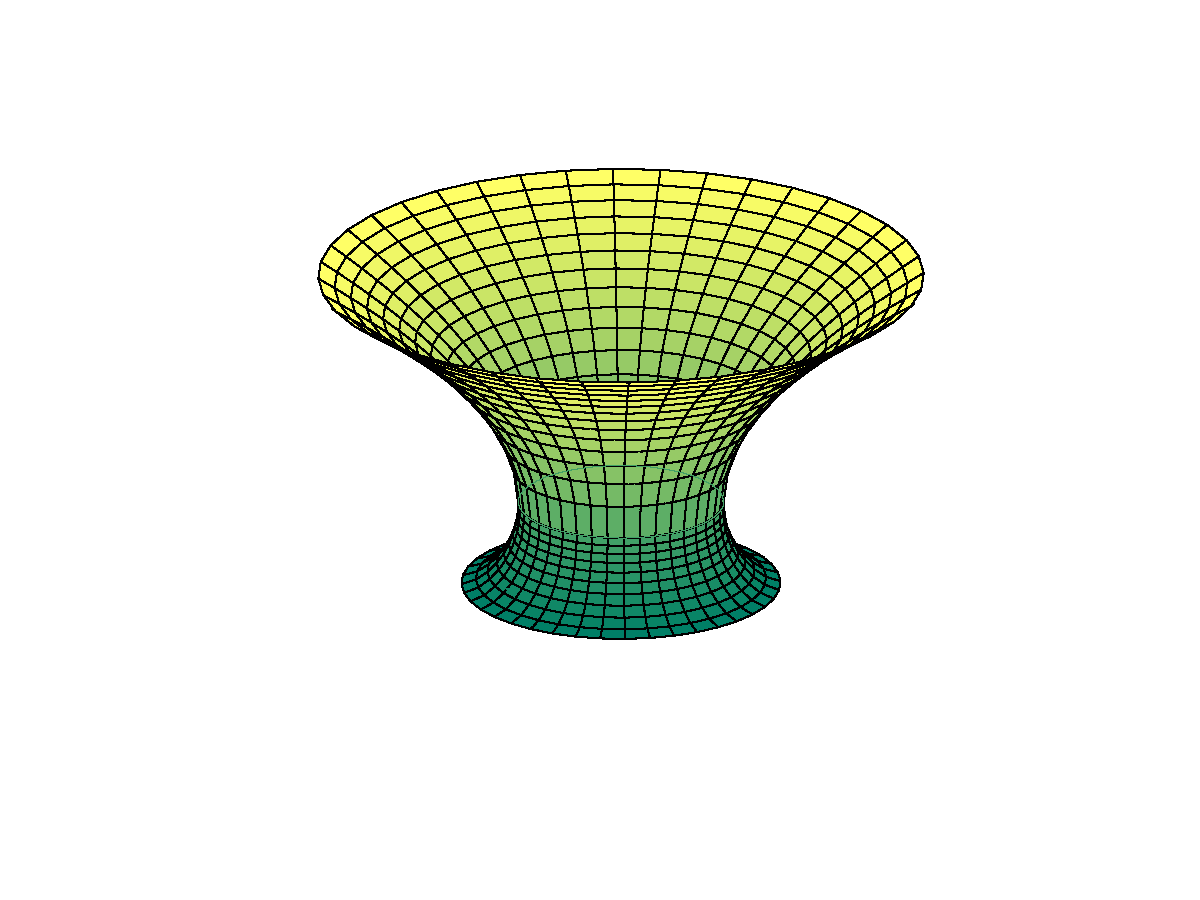}
    \caption{The embedding diagram of BE wormhole with $r_0=1$ and different masses $M$: $M=-1$ (left); $M=0$ (middle); $M=1$ (right)}
    \label{WH_Embedding}
\end{figure}

If we set $\psi=0$, then the analytical form of static and spherically symmetric BE wormhole is given by \cite{Blazquez-Salcedo:2018ipc}
\begin{align}
   \phi(\eta) &=  \frac{Q}{\eta_0 } \left( \arctan  \frac{\eta}{\eta_0}  - \frac{\pi}{2} \right) \,, \label{solp} \\
    ds^2 &= - e^{F(\eta)} dt^2 + e^{-F(\eta)} \left[ d\eta^2 + \left(  \eta^2 + \eta^2_0 \right) \left( d\theta^2 + \sin^2 \theta d\varphi^2 \right) \right] \,,
\end{align}
where  
\begin{equation}
    F(\eta) =  \frac{2M}{\eta_0 } \left( \arctan  \frac{\eta}{\eta_0}  - \frac{\pi}{2} \right) \,, \label{solF} 
\end{equation}
with $\eta_0$ is the throat parameter, $M$ is the Arnowitt-Deser-Misner (ADM) mass of wormhole, $Q$ is the scalar charge of $\phi(\eta)$. These parameters are constrained by the simple relation $M^2 + \eta^2_0 = Q^2$. The range of radial coordinate $\eta$ is $-\infty < \eta < \infty$; hence the BE wormhole possesses two asymptotically flat regions when $\eta \rightarrow \pm \infty$. It is obvious $F(\eta) \rightarrow 1$ and $\phi(\eta) \rightarrow 0$ as $\eta \rightarrow \infty$. However, we can reparametrize the coordinate as $\bar{t} = e^{\pi M/\eta_0} t$, $\bar{\eta} = e^{-\pi M/\eta_0} \eta$, so that the metric function $F(\eta)$ becomes asymptotically flat when $\eta \rightarrow -\infty$. As we observe from the solution of the BE wormhole, we could scale all the parameters by $\eta_0$. 

The geometry of the BE wormhole is relatively simple; first, we can define its circumferential radius on the equatorial plane $(\theta=\pi/2)$ as
\begin{equation}
    R_e = \sqrt{g_{\varphi \varphi}} \bigg|_{\theta=\pi/2} = e^{-f/2} \sqrt{\eta^2+\eta^2_0} \,.
\end{equation}
Then, we inspect that $R_e$ contains a global minimum, which indicates it only possesses a single throat at the radial coordinate $\eta = 2M$. The shape of the wormhole throat can be visualized in Fig.~\ref{WH_Embedding} by embedding the throat on the equatorial plane to the three-dimensional cylindrical coordinate $(z,\rho,\varphi)$ in the Euclidean space, 
\begin{equation}
     e^{-f} d\eta^2 + e^{-f} (\eta^2 + \eta^2_0) d\varphi^2 = dz^2 +  d\rho^2 + \rho^2 d\varphi^2 \,,
\end{equation}
which yields
\begin{equation}
    z = \pm \int \sqrt{ e^{-f} - \left( \frac{d}{d\eta}  e^{-f} (\eta^2 + \eta^2_0) \right)^2 } d\eta \,, 
\end{equation}
where $+$ sign denotes the upper part of the wormhole and $-$ sign denotes the lower part of the wormhole. The structure of the BE wormhole is symmetric for the massless case $(M=0)$ but asymmetric for the massive case $(M \neq 0)$.

\subsection{\label{sec:fieldeqns}Double-null formalism}

The most generic expression for a spherically symmetric metric is employed as an ansatz in the double-null formalism:
\begin{eqnarray}
ds^{2} = -\alpha^{2}(u,v) du dv + r^{2}(u,v) d\Omega^{2} \,,
\end{eqnarray}
where the embedded coordinate is $[u, v, \theta, \varphi]$ with $u$ and $v$ denote ingoing and outgoing null directions, respectively. 
The direct substitution of the above metric into the equations of motion, Eqs.~\eqref{eom1}-\eqref{eom3} can yield a set of second-order PDEs, thus we can reduce second-order derivative of functions as first-order derivatives of functions for the sake of numerical computation by redefining several functions as $\alpha_{,u}/\alpha \equiv h$, $\alpha_{,v}/\alpha \equiv d$, $r_{,u} \equiv f$, $r_{,v} \equiv g$, $\sqrt{4\pi} \psi \equiv s$, $s_{,u} \equiv w$, $s_{,v} \equiv z$, $\sqrt{4\pi} \phi \equiv S$, $S_{,u} \equiv W$, and $S_{,v} \equiv Z$. Then,  the non-vanishing components of the Einstein tensor $G_{\mu\nu}$ obtained as follows \cite{Nakonieczna:2018tih}:
\begin{eqnarray}
&& G_{uu} = -\frac{2}{r} \left(f_{,u}-2fh \right) \,,\\
&& G_{uv} = \frac{1}{2r^{2}} \left( 4 rf_{,v} + \alpha^{2} + 4fg \right) \,,\\
&& G_{vv} = -\frac{2}{r} \left(g_{,v}-2gd \right) \,,\\
&& G_{\theta\theta} = -4\frac{r^{2}}{\alpha^{2}} \left(d_{,u}+\frac{f_{,v}}{r}\right) \,,
\end{eqnarray}
and the components of the energy-momentum tensor $T_{\mu\nu}$:
\begin{eqnarray}
&& T_{uu} = \frac{1}{4\pi} \left( w^{2} - W^{2} \right), \\
&& T_{uv} = 0, \\
&& T_{vv} = \frac{1}{4\pi} \left( z^{2} - Z^{2} \right), \\
&& T_{\theta\theta} = \frac{r^{2}}{2\pi\alpha^{2}} \left( wz - WZ \right).
\end{eqnarray}
Note that $G_{\theta\theta} = \sin^{-2}\theta G_{\varphi\varphi}$ due to the spherical symmetry, hence we do not need to consider the $\varphi\varphi$-component in the computation.

We obtain the following PDEs by algebraically manipulating $G_{\mu \nu}=T_{\mu \nu}$:
\begin{eqnarray}
\label{eq:E1}f_{,u} &=& 2 f h - r \left( w^{2} - W^{2} \right),\\
\label{eq:E2}g_{,v} &=& 2 g d - r \left( z^{2} - Z^{2} \right),\\
\label{eq:E3}f_{,v}=g_{,u} &=& -\frac{\alpha^{2}}{4r} - \frac{fg}{r} ,\\
\label{eq:E4}h_{,v}=d_{,u} &=& - \left( wz - WZ \right) - \frac{f_{,v}}{r},
\end{eqnarray}
and the two Klein-Gordon equations as:
\begin{eqnarray}
\label{eq:s}z_{,u} = w_{,v} &=& - \frac{fz}{r} - \frac{gw}{r},\\
\label{eq:S}Z_{,u} = W_{,v} &=& - \frac{fZ}{r} - \frac{gW}{r}.
\end{eqnarray}

We solve Eqs.~(\ref{eq:E3}), (\ref{eq:E4}), (\ref{eq:s}), and (\ref{eq:S}) for the functions $r$, $\alpha$, $\phi$, and $\psi$, respectively. Thus, the remaining equations, Eqs.~(\ref{eq:E1}) and (\ref{eq:E2}) are the constraint equations, which can be assigned as the boundary conditions in the numerical computation.

\subsection{Boundary conditions}

To solve the PDEs, we need to impose initial conditions for functions $\alpha, h, d, r, f, g, s, w, z, S, W, Z$ on the initial surfaces $u=u_{\mathrm{i}}$ and $v=v_{\mathrm{i}}$. We can choose $u_{\mathrm{i}}=v_{\mathrm{i}}=0$ for simplicity. Besides, a direct comparison of the metric of static BE wormhole with double-null coordinate yields
\begin{eqnarray}
r(u,v) = \sqrt{R_e}  \,.
\end{eqnarray}

First, we impose the analytical solution of BE wormhole on the initial hypersurfaces $(u,0)$ and $(0,v)$ in the double-null coordinates as:
\begin{eqnarray}
\phi(\eta) &=& \frac{\sqrt{M^{2} + \eta_{0}^{2}}}{\eta_{0}} \left( \arctan \frac{\eta}{\eta_{0}} - \frac{\pi}{2} \right), \label{eq:scalar}\\
r(\eta) &=& \sqrt{\eta^{2} + \eta_{0}^{2}} \exp \left[ - \frac{M}{\eta_{0}} \left( \arctan \frac{\eta}{\eta_{0}} - \frac{\pi}{2} \right) \right].
\end{eqnarray}
For simplicity, we also assume a gauge choice of $\eta$ as a simple combination of $u$ and $v$ as:
\begin{eqnarray}
\eta = \frac{u - v + 2M}{2}.
\end{eqnarray}
Thus, this allows us to set the analytical solution of BE wormhole as boundary conditions for $f(u,0)$, $s(u,0)$, and $w(u,0)$ on the ingoing null surface and $g(0,v)$, $s(0,v)$, and $z(0,v)$ on the outgoing null surface. (Regarding the assignment of boundary condition for $\alpha(0,0)$, see also \cite{Hansen:2009kn,Hwang:2011mn,Hwang:2011mn}):
\begin{eqnarray}
 f(u,0) = \left.r_{,u}\right|_{v=0} &=& \frac{u/2}{\left( 4\eta_{0}^{2} + \left(u + 2M\right)^{2} \right)^{1/2}} \exp \left[ - \frac{M}{\eta_{0}} \left( \arctan \frac{u + 2M}{2\eta_{0}} - \frac{\pi}{2} \right) \right],\\
f_{,u}(u,0) = \left.r_{,uu}\right|_{v=0} &=& \frac{2 \left( M^{2} + \eta_{0}^{2} \right)}{\left( 4\eta_{0}^{2} + \left(u + 2M\right)^{2} \right)^{3/2}} \exp \left[ - \frac{M}{\eta_{0}} \left( \arctan \frac{u + 2M}{2\eta_{0}} - \frac{\pi}{2} \right) \right],\\
g(0,v) = \left.r_{,v}\right|_{u=0} &=& \frac{v/2}{\left( 4\eta_{0}^{2} + \left(-v + 2M\right)^{2} \right)^{1/2}} \exp \left[ - \frac{M}{\eta_{0}} \left( \arctan \frac{-v + 2M}{2\eta_{0}} - \frac{\pi}{2} \right) \right],\\
g_{,v}(0,v)=\left.r_{,vv}\right|_{u=0} &=& \frac{2 \left( M^{2} + \eta_{0}^{2} \right)}{\left( 4\eta_{0}^{2} + \left(-v + 2M\right)^{2} \right)^{3/2}} \exp \left[ - \frac{M}{\eta_{0}} \left( \arctan \frac{-v + 2M}{2\eta_{0}} - \frac{\pi}{2} \right) \right],\\
g_{,u}(0,0)= f_{,v}(0,0)= \left.r_{,uv}\right|_{u=v=0} &=& - \frac{1}{4 \sqrt{M^{2} +\eta_{0}^{2}}} \exp \left[ \frac{M}{\eta_{0}} \left( - \arctan \frac{M}{\eta_{0}} + \frac{\pi}{2} \right) \right].
\end{eqnarray}
The boundary condition for $\alpha(0,0)$ is derived from Eq. (\ref{eq:E3}) with  $r_{,u}(0,0) = r_{,v}(0,0) = 0$, thus we obtain $\alpha^{2}(0,0) = -4 r(0,0) r_{,uv}(0,0)$ as
\begin{eqnarray}
\alpha(0,0) = \exp \left[ \frac{M}{\eta_{0}} \left( -\arctan \frac{M}{\eta_{0}} + \frac{\pi}{2} \right) \right].
\end{eqnarray}

By having the above sufficient information, we can determine the boundary conditions for the remaining functions as follows:
\begin{description}
\item[Ingoing null surface:]
$S(u,0)$ and $W(u,0)$ can be chosen by hand. By plugging them into Eq.~(\ref{eq:E1}), we obtain $h(u,0) = 0$. 
We obtain $d(u,0)$ from Eq.~(\ref{eq:E4}), $g(u,0)$ from Eq.~(\ref{eq:E3}), $z(u,0)$ from Eq.~(\ref{eq:s}), and $Z(u,0)$ from Eq.~(\ref{eq:S}).
\item[Outgoing null surface:] 
$S(0,v)$ and $W(0,v)$ can be chosen by hand. By plugging them into Eq.~(\ref{eq:E2}), we obtain $d(0,v) = 0$. 
We obtain $h(0,v)$ from Eq.~(\ref{eq:E4}), $f(0,v)$ from Eq.~(\ref{eq:E3}), $w(0,v)$ from Eq.~(\ref{eq:s}), and $W(0,v)$ from Eq.~(\ref{eq:S}).
\end{description}

After solving the PDEs numerically, we check the convergence of our numerical results and present them in Appendix A.

\subsection{Free parameters in the simulation}

We briefly describe the tuning of some parameters in our simulation. The BE wormhole can be characterized by two parameters, which are $M$ and $\eta_0$ from the constraint $M^2+\eta^2_0=Q^2$. It can be conveniently described by only one parameter $M$ if we scale $M$ by $\eta_0$ as $M/\eta_0$. In the simulation, the collapse of BE wormhole is triggered by the propagation of pulses in the form of normal scalar $Z(0,v)$ and phantom $z(0,v)$ fields in the spacetime, then we introduce their analytical expressions along the outgoing null direction ($u = 0$) for the range $ v_{\mathrm{i}} \leq v \leq v_{\mathrm{f}}$:
\begin{eqnarray}
z(0,v) = - \frac{\pi}{2} \frac{s^{v}_{0}}{v_{\mathrm{f}} - v_{\mathrm{i}}} \sin{ \left( \pi \frac{v - v_{\mathrm{i}}}{v_{\mathrm{f}} - v_{\mathrm{i}}} \right)}, \\
Z(0,v) = - \frac{\pi}{2} \frac{S^{v}_{0}}{v_{\mathrm{f}} - v_{\mathrm{i}}} \sin{ \left( \pi \frac{v - v_{\mathrm{i}}}{v_{\mathrm{f}} - v_{\mathrm{i}}} \right)},
\end{eqnarray}
and along the ingoing null direction ($v = 0$) for $u_{\mathrm{i}} \leq u \leq u_{\mathrm{f}}$:
\begin{eqnarray}
w(u,0) = - \frac{\pi}{2} \frac{s^{u}_{0}}{u_{\mathrm{f}} - u_{\mathrm{i}}} \sin{ \left( \pi \frac{u - u_{\mathrm{i}}}{u_{\mathrm{f}} - u_{\mathrm{i}}} \right)}, \\
W(u,0) = - \frac{\pi}{2} \frac{S^{u}_{0}}{u_{\mathrm{f}} - u_{\mathrm{i}}} \sin{ \left( \pi \frac{u - u_{\mathrm{i}}}{u_{\mathrm{f}} - u_{\mathrm{i}}} \right)},
\end{eqnarray}
We explain in detail the parameters in the above expressions:
\begin{itemize}
\item[--] \textit{Amplitudes of pulses}: ingoing scalar field $s^{v}_{0}$, ingoing phantom field $S^{v}_{0}$, outgoing scalar field $s^{u}_{0}$, and outgoing phantom field $S^{u}_{0}$.
\item[--] \textit{Time interval of pulses}: ingoing pulse only exists in $v_{\mathrm{i}} \leq v \leq v_{\mathrm{f}}$; outgoing pulse only exists in $u_{\mathrm{i}} \leq u \leq u_{\mathrm{f}}$.
\end{itemize}

\begin{figure}[h]
\centering
 \includegraphics[scale=0.2]{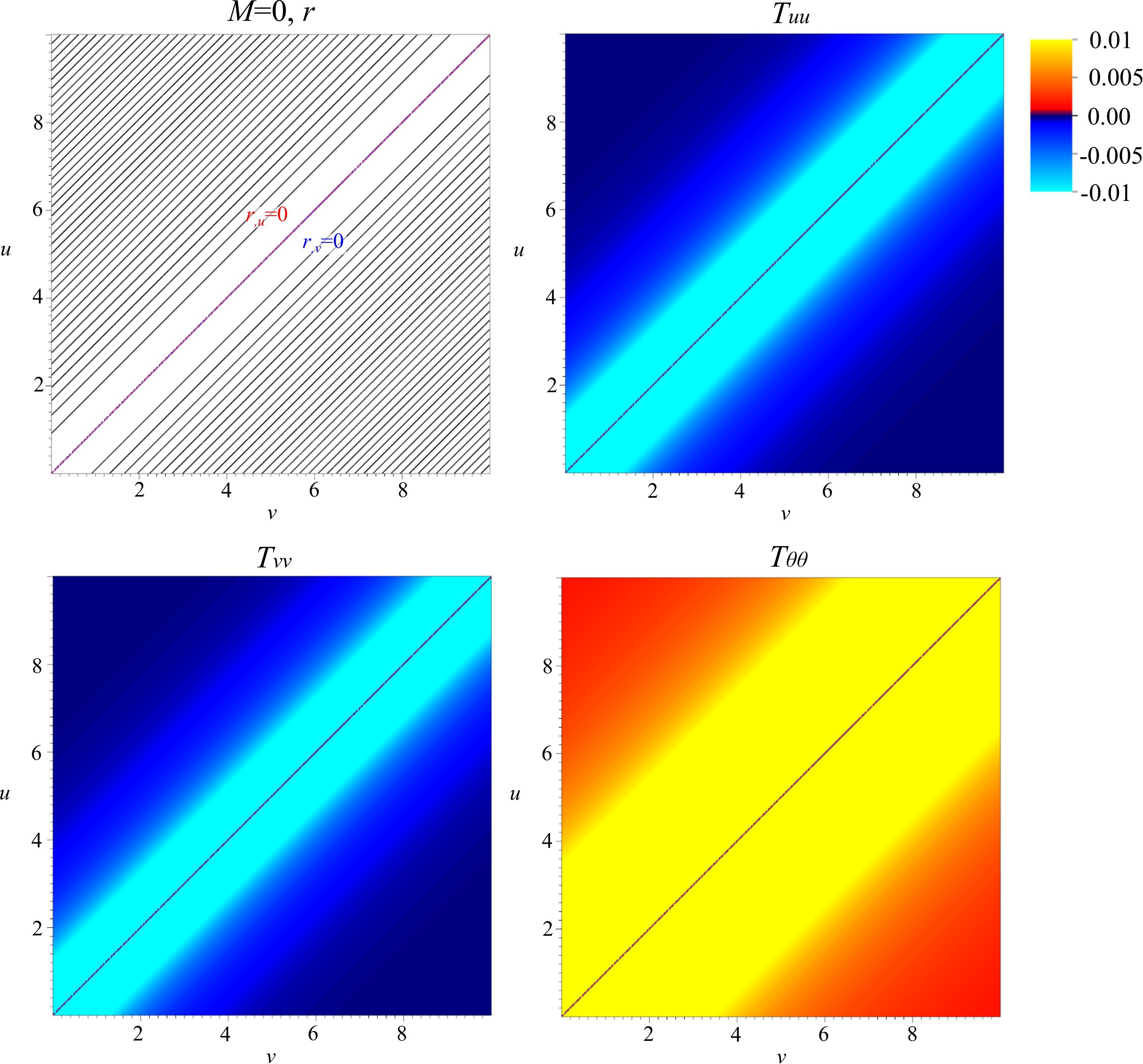}
\caption{Some characteristics of massless BE wormhole $(M=0)$ in the double-null coordinates (clockwise direction): $r(u,v)$, $T_{uu}(u,v)$, $T_{vv}(u,v)$, and $T_{\theta\theta}(u,v)$. $r_{,v} = 0$ (blue) and $r_{,u} = 0$ (red) represent apparent horizons for an outgoing and an ingoing null direction, respectively. The overlapping of $r_{,v} = r_{,u}  = 0$ denotes the wormhole throat. The negative values of $T_{uu}$ and $T_{vv}$ are described by the color gradient of blue, but the positive values of $T_{\theta \theta}$ are described by the color gradient of red and yellow.}
\label{fig:M_nopulse}
\end{figure}

\section{\label{sec:res}Dynamics of BE wormhole}

There are two main parts in this section. The first part introduces some properties of the static BE wormhole in the double-null coordinates. Then, we study the gravitational collapse of the BE wormhole, which is triggered by the emissions of normal scalar and phantom pulses in the second main part.

\subsection{Static solution (without pulse)}

\subsubsection{Massless limit}

Fig.~\ref{fig:M_nopulse} illustrates the main features of the BE wormhole for the massless case $M = 0$, where the spacetime exhibits symmetry about $\eta = 0$. The contour plot of $r(u,v)$ depicts curves of constant areal radius $r$, with the wormhole throat identified by the overlap of the apparent horizons $r_{,v} = 0$ and $r_{,u} = 0$ \cite{Ashtekar:2004cn}. The throat forms approximately $45^\circ$ angle with the outgoing horizontal $(u=\mathrm{constant},v)$ and ingoing vertical $(u,v=\mathrm{constant})$ null directions, divides the whole wormhole spacetime into two identical structures, analogous to the middle figure of Fig.~\ref{WH_Embedding}. Note that the areal radius  $r$ increases monotonically when moving away from the wormhole throat to the lower or upper part of the wormhole. Thus, an ingoing observer (constant $v$) can observe $r$ decreases to a minimum value at the throat when moving from past infinity toward the throat, and then $r$ increases monotonically when crossing the throat, since $r_{,u}$ changes its sign at the throat, marking the transition of wormhole geometry from contraction to expansion. An outgoing observer (constant $u$) also observes a similar effect when crossing from the past infinity to the future infinity where $r_{,v}$ changes its sign at the throat.

The existence of BE wormhole requires the violation of energy conditions; thus, it would be interesting to analyze its energy-momentum tensor components: $T_{uu}$, $T_{vv}$, and $T_{\theta\theta}$. Here, $T_{uu}$ quantifies the ingoing energy flux as measured by an outgoing observer, while $T_{vv}$ corresponds to the outgoing energy flux as seen by an ingoing observer. The null energy condition (NEC) holds only if both $T_{uu}$ and $T_{vv}$ are positive. However, Fig.~\ref{fig:M_nopulse} demonstrates that both components are negative throughout the spacetime, confirming the violation of NEC for all ingoing and outgoing observers. The color gradient further reveals that this violation is maximum at the wormhole throat, where the phantom field is most concentrated, and gradually diminishes to zero at infinity.

\subsubsection{Massive Case}


\begin{figure}[h]
\centering
 \includegraphics[scale=0.2]{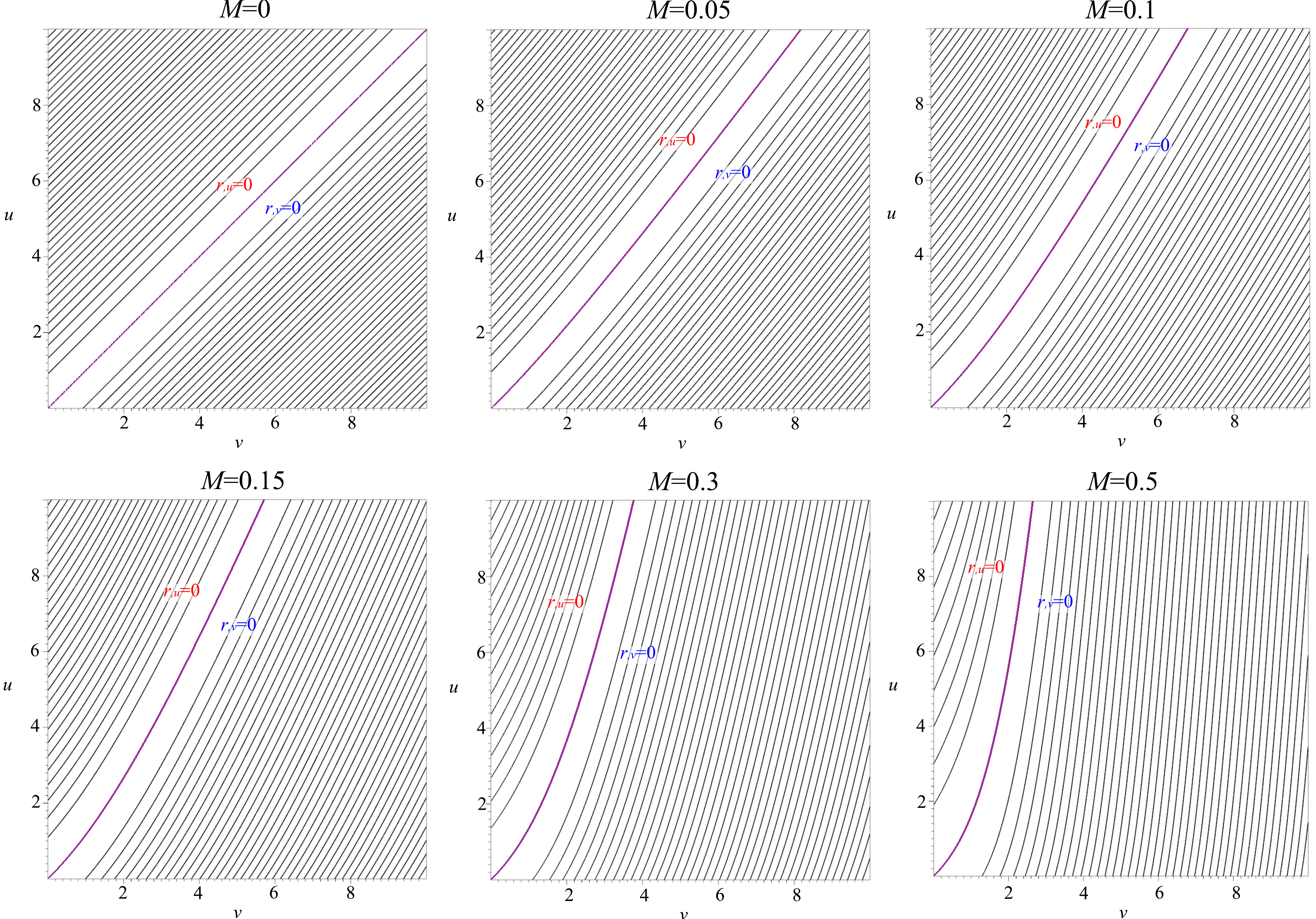}
\caption{The contour plot of $r(u,v)$ for massive BE wormhole by varying $M = 0$, $0.05$, $0.1$, $0.15$, $0.3$, and $0.5$.}
\label{fig:M_nopulse_varyingM}
\end{figure}

Fig.~\ref{fig:M_nopulse_varyingM} exhibits the contour plot of $r(u,v)$ for the BE wormhole with varying $M$. As $M$ increases, the throat gradually deviates from the $45^\circ$ alignment, as observed in the massless case, causing the wormhole's upper and lower parts to be no longer identical. Such deviation can also be analogously seen in the embedding diagrams (Fig.~\ref{WH_Embedding}), where the wormhole structure is asymmetric when $M>0$. However, it is possible to restore the $45^\circ$ alignment of curves $r(u,v)$ with the ingoing and outgoing null directions for a massive BE wormhole since there is a gauge freedom to reparametrize the wormhole solution.

\begin{figure}[h]
\centering
 \includegraphics[scale=0.2]{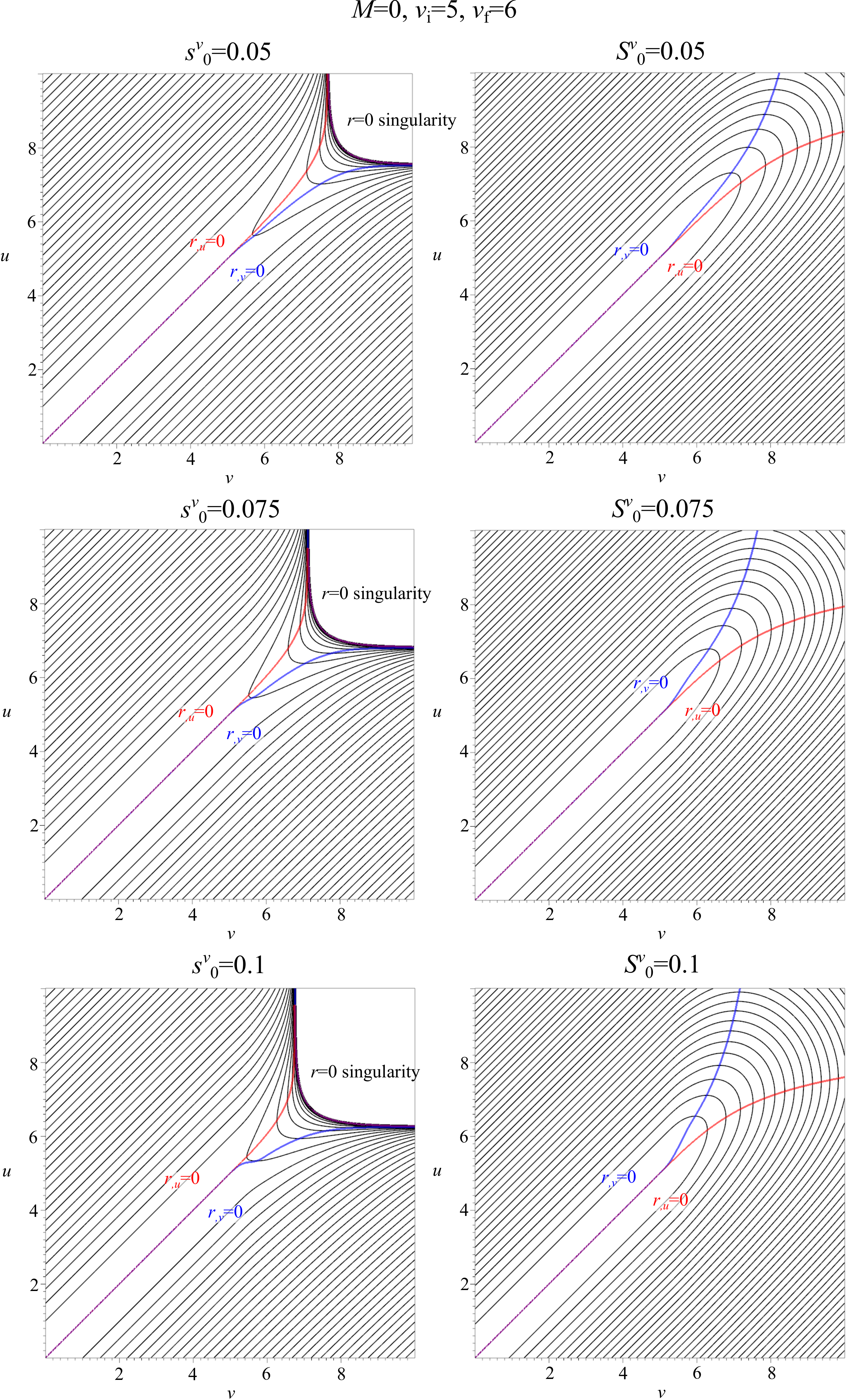}
\caption{
Destabilization of the massless BE wormhole $(M = 0)$ induced by a single pulse along the ingoing null direction from $v_i=5$ to $v_f=6$. The left and right figures correspond to the pulses of normal scalar $s^{v}_{0}$ and phantom fields $S^{v}_{0}$ with different amplitudes, respectively.
}
\label{fig:M=0_singlepulse}
\end{figure}

\begin{figure}[h]
\centering
 \includegraphics[scale=0.2]{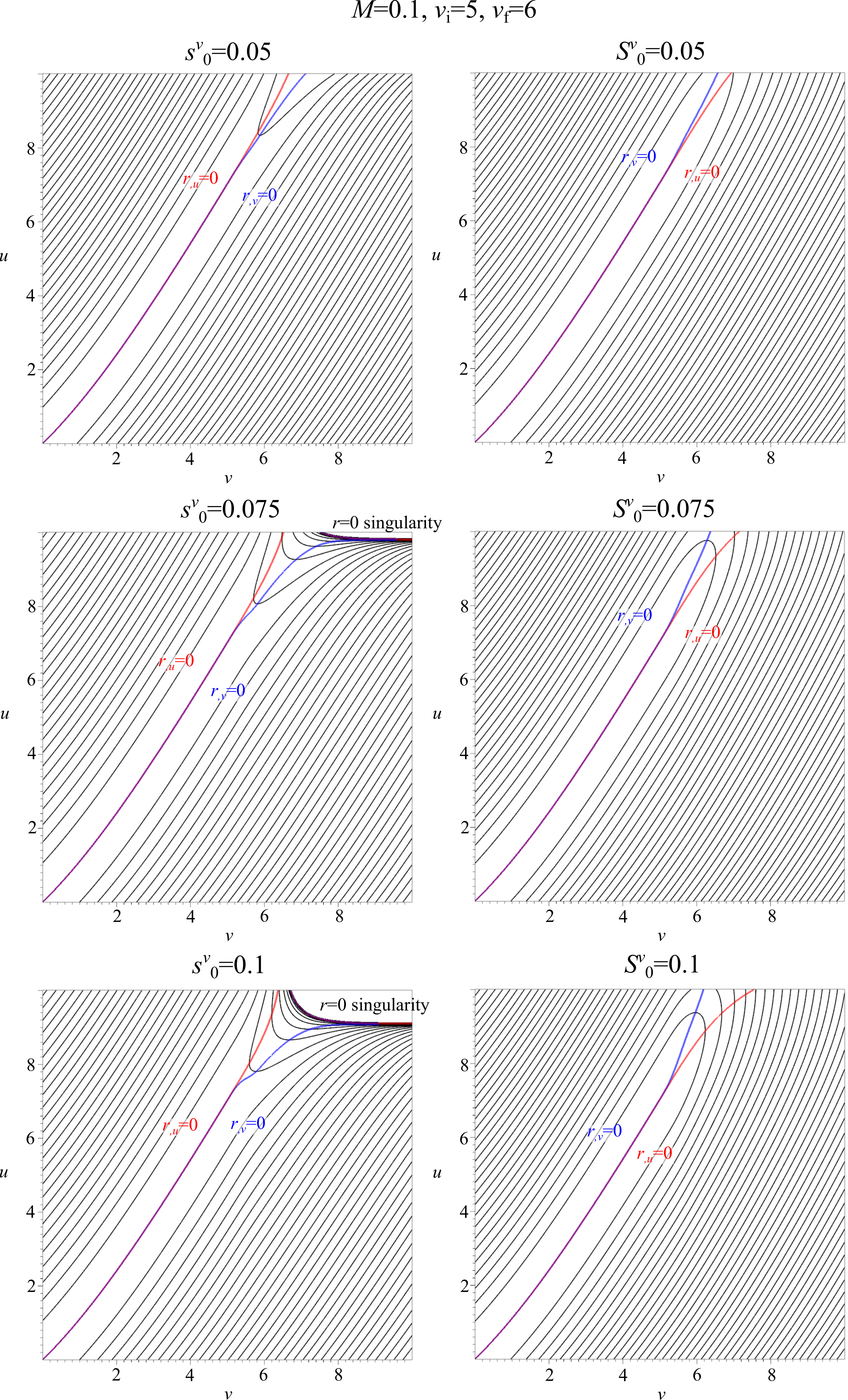}
\caption{
Destabilization of the massive BE wormhole with $M = 0.1$ induced by a single pulse along the ingoing null direction from $v_i=5$ to $v_f=6$. The left and right figures correspond to the pulses of normal scalar $s^{v}_{0}$ and phantom fields $S^{v}_{0}$ with different amplitudes, respectively.
}
\label{fig:M=0.1_singlepulse}
\end{figure}

\begin{figure}[h]
\centering
 \includegraphics[scale=0.3]{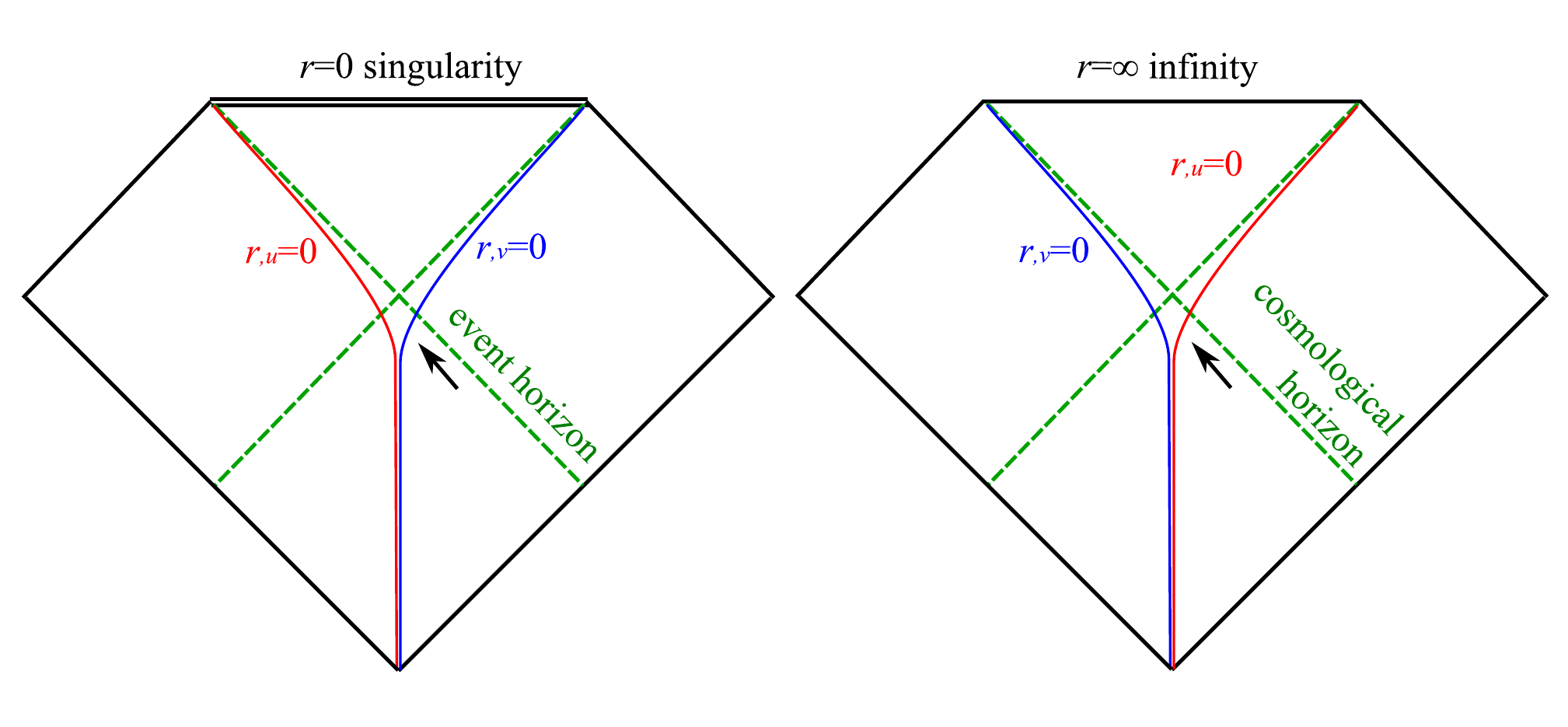}
\caption{
The Penrose diagrams illustrate the causal structure of the BE wormhole, which is destabilized by a single pulse along the ingoing null direction (black arrow) of a normal scalar field (left) and a phantom field (right). Separation of the apparent horizons $r_{,u} = 0$ (red) and $r_{,v} = 0$ (blue) leads to two distinct outcomes: for the normal field, an event horizon (green dashed) forms, shielding a $r = 0$ spacelike singularity; for the phantom field, a cosmological horizon (green dashed) terminates at future null infinity $r = \infty$, avoiding singularity formation.
}
\label{fig:concep}
\end{figure}


\subsection{Destabilization of wormhole}

\subsubsection{By a single pulse}

We investigate the stability of BE wormhole under the influence of ingoing normal scalar and phantom pulses along the outgoing null hypersurface $v$, as demonstrated numerically in Figs.~\ref{fig:M=0_singlepulse} and \ref{fig:M=0.1_singlepulse}. For the $M = 0$ case (Fig.~\ref{fig:M=0_singlepulse}), the left and right panel figures depict normal scalar and phantom pulses with varying amplitudes, respectively.

In the normal scalar pulse scenario, the curves $r_{,v} = 0$ and $r_{,u} = 0$, which initially coincide at the wormhole throat to form an apparent horizon, diverge when a pulse of normal scalar field with the amplitude $s^v_0=0.05$ is emitted from $(u=0, v_i=5)$ and propagates toward the apparent horizon. This splits the throat into two distinct apparent horizons $r_{,v} = 0$ and $r_{,u} = 0$, destabilizing the wormhole geometry. Consequently, a curvature singularity forms at $r = 0$, hidden by these horizons with $r_{,v} = 0$ and $r_{,u} = 0$. Observers located in the lower asymptotically flat region (right side) detect $r_{,v} = 0$ horizon, while observers in the upper region (left side) detect $r_{,u} = 0$. Both independently indicate the collapse of the BE wormhole into a Schwarzschild black hole, as the throat no longer connects the two asymptotic regions (see left of Fig.~\ref{fig:concep}). As we increase the amplitudes of pulses $s^v_0$ to $0.075$ and $0.1$, the singularity appears relatively earlier, and this is related to the fact that we increase the energy of the pulse, which can accelerate the process. Besides, two observers on opposite sides of the wormhole can communicate with each other if they stay outside the event horizon, as shown in Fig.~\ref{fig:concep} (left).

In the phantom pulse scenario, the curves $r_{,v} = 0$ and $r_{,u} = 0$ initially coincide at the wormhole throat, forming the apparent horizon. Unlike the normal scalar case, no black hole and curvature singularity form when a phantom pulse with the amplitude $S^v_0=0.05$ is emitted from $(u=0, v_i=5)$ and propagates toward the apparent horizon, even though the apparent horizon splits into distinct $r_{,v} = 0$ and $r_{,u} = 0$ curves. Observers in the lower region of the wormhole (right side) detect $r_{,u} = 0$ horizon, while those in the upper region (left side) observe $r_{,v} = 0$. These distinct horizons show inflationary dynamics in the BE wormhole, which ultimately decouples the two asymptotic regions by the cosmological horizon with $r_{,u} = 0$ and $r_{,v} = 0$ (see right of Fig.~\ref{fig:concep}). Consequently, the BE wormhole retains traversability only temporarily immediately after the phantom pulse as the spacetime geometry begins to separate. The inflationary expansion irreversibly destroys the throat as time passes, permanently cutting the connection.

Fig.~\ref{fig:M=0.1_singlepulse} demonstrates the destabilization of the massive BE wormhole with $M=0.1$ via a single-pulse emission of an ingoing normal scalar field (left panel) and phantom field (right panel) for $v_i=5$ to $v_f=6$. Increasing the amplitude $s^v_0$ of the normal scalar field accelerates the formation of a black hole with an event horizon and a spacelike singularity $r=0$. Meanwhile, increasing the amplitude $S^v_0$ of the phantom field widens the separation between the apparent horizons $r_{,u}=0$ and $r_{,v}=0$. Hence, the stability of the massive wormhole cannot be enhanced, and it becomes worse in this scenario.

\subsubsection{By collision of two pulses at the wormhole throat}

Previously, we observed that a single pulse destabilizes the BE wormhole, triggering either black hole formation for a normal scalar pulse or permanent separation of the asymptotic regions for a phantom pulse. Here, we investigate scenarios where two pulses collide with perfect timing at the wormhole throat. By adjusting the parameters of the pulses, we explore whether their interaction could counteract the destabilizing effects as observed in single-pulse cases. Specifically, can fine-tuned collisions stabilize the throat, delaying or preventing horizon formation and maintaining traversability?

\begin{figure}
\centering
 \includegraphics[scale=0.2]{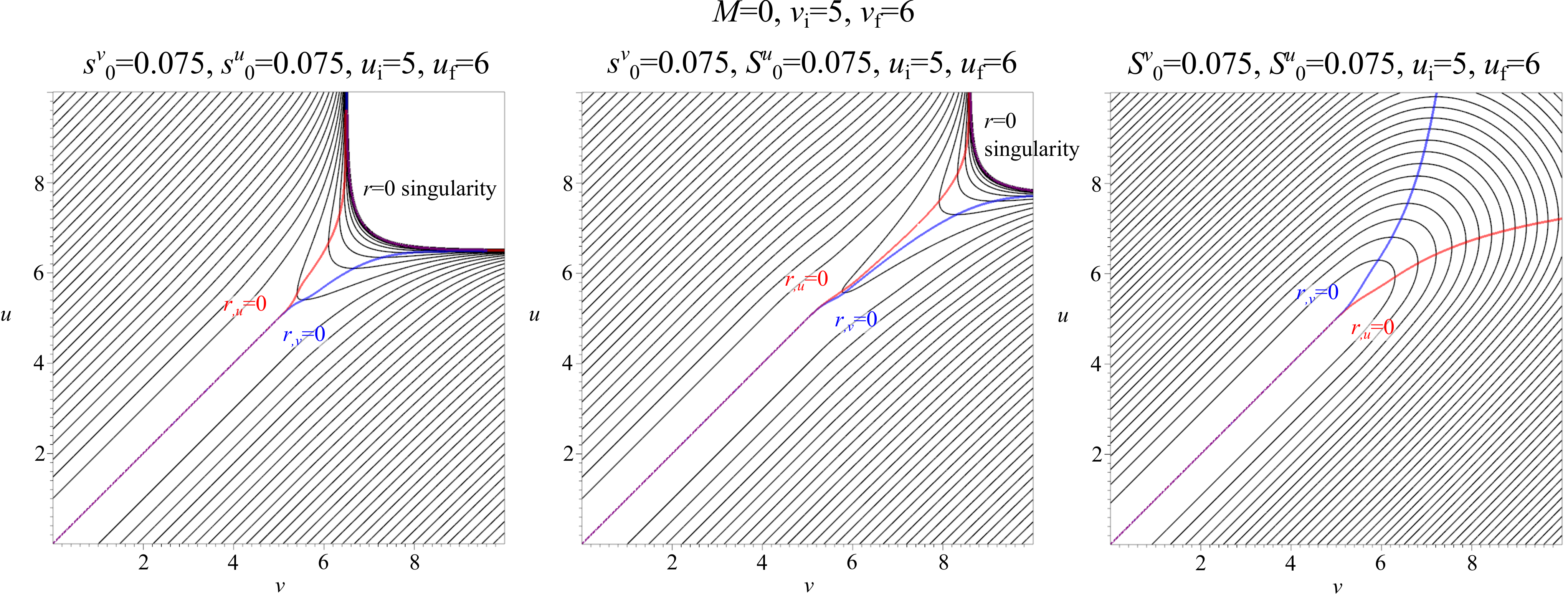}
\caption{
Destabilization of the massless BE wormhole $M = 0$ via collisions of two pulses with equal magnitude of amplitude at the throat. Pulses are emitted along ingoing ($v_i=5$ to $v_f=6$) and outgoing ($u_i=5$ to $u_f=6$) null directions with amplitudes $(s^v_0, S^v_0)$ and $(s^u_0, S^u_0)$, respectively, for three cases: (left) two normal scalar fields $(s^v_0, s^u_0)$; (middle) normal scalar $s^v_0$ and phantom fields $S^u_0$; (right) two phantom fields $(S^v_0, S^u_0)$.
}
\label{fig:M=0_twopulses}
\end{figure}
\begin{figure}[h]
\centering
 \includegraphics[scale=0.2]{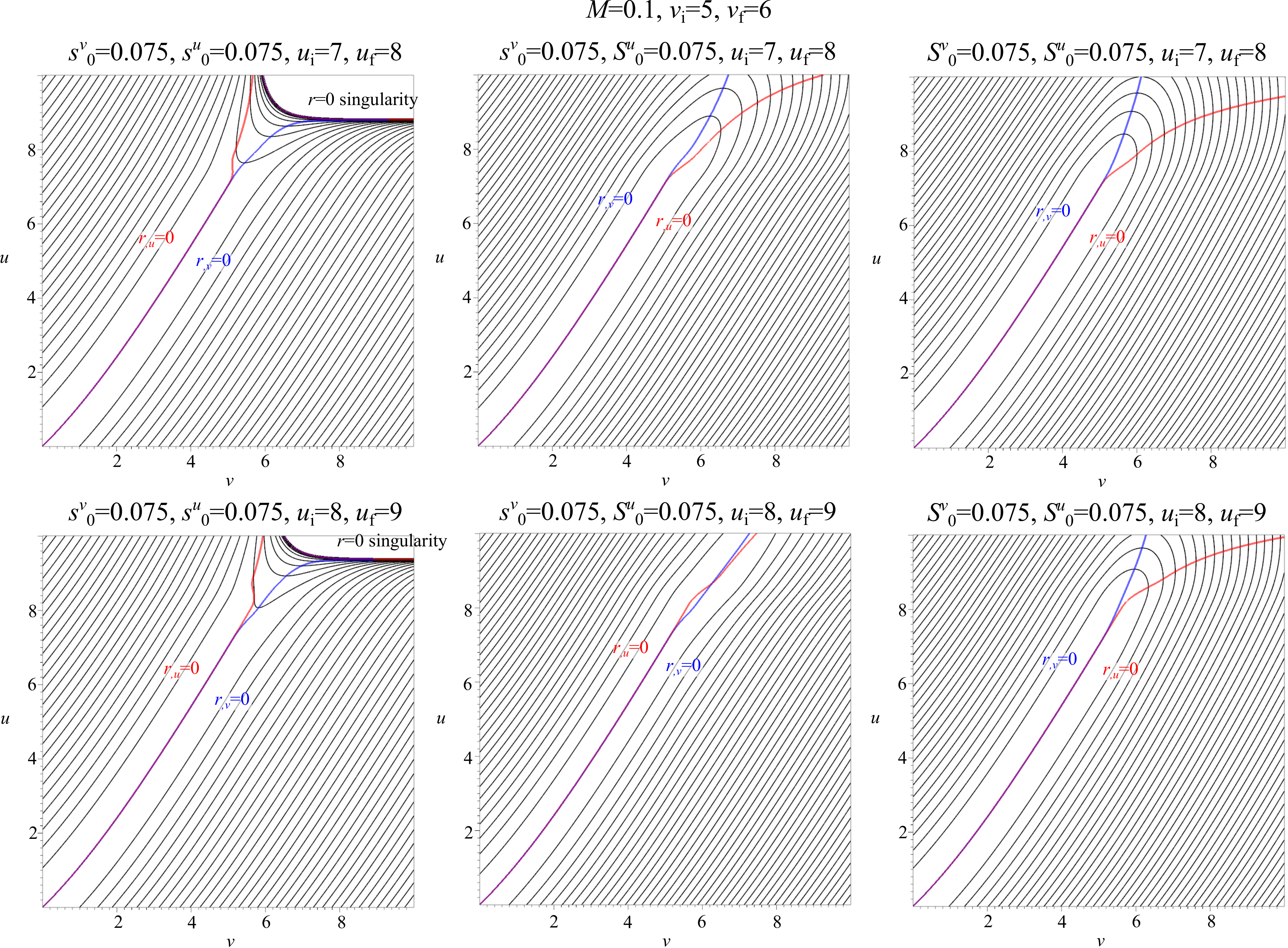}
\caption{
Destabilization of the massive BE wormhole $M = 0.1$ via collisions of two pulses with equal magnitude of amplitude at the throat. Pulses are emitted along ingoing ($v_i=5$ to $v_f=6$) and outgoing (upper: $u_i=7$ to $u_f=8$; lower: $u_i=8$ to $u_f=9$) null directions with amplitudes $(s^v_0, S^v_0)$ and $(s^u_0, S^u_0)$, respectively, for three panels: (left) two normal scalar fields $(s^v_0, s^u_0)$; (middle) normal scalar $s^v_0$ and phantom fields $S^u_0$; (right) two phantom fields $(S^v_0, S^u_0)$.
}
\label{fig:M_twopulses}
\end{figure}

Fig.~\ref{fig:M=0_twopulses} illustrates collisions of two scalar pulses in combinations of normal scalar and phantom fields at the throat of a massless BE wormhole $M=0$. In this framework, $s_0^v$ and $s_0^u$ denote normal scalar pulses propagating in the ingoing $v$-direction and outgoing $u$-direction, respectively, while $S_0^v$ and $S_0^u$ represent phantom scalar pulses in the same directions. When both ingoing and outgoing pulses are normal scalar fields (left panel), the collision forms a Schwarzschild black hole and a curvature singularity at $r=0$, replicating the dynamics observed in single-pulse scenarios. Conversely, collisions of two phantom pulses (right panel) trigger inflationary expansion of spacetime, decoupling two asymptotic regions by cosmological horizons, also analogous to the single phantom pulse case.

A notable feature arises in the middle panel of Fig.~\ref{fig:M=0_twopulses}, which depicts a collision at the throat of two pulses with identical magnitude and timing but opposite types: an ingoing normal scalar pulse $s_0^v$ and an outgoing phantom scalar pulse $S_0^u$. We might naively expect that this collision can prevent the formation of a (Schwarzschild) black hole and curvature singularity in the wormhole spacetime, but they eventually exist. However, the curvature singularity forms later where the wormhole spacetime can be maintained slightly longer than the purely normal-pulse collision (left panel of Fig.~\ref{fig:M=0_twopulses}). This indicates that the collision of mixed pulses does not stabilize the wormhole; it just temporarily extends the traversability of the wormhole by delaying singularity formation. Thus, the wormhole geometry remains unstable in this hybrid scenario, but its lifetime is prolonged compared to other cases in Figs.~\ref{fig:M=0_twopulses} and \ref{fig:M_twopulses}.

The dynamics of pulse collisions can be extended to BE wormhole with non-vanishing mass $M \neq 0$, as shown in Fig.~\ref{fig:M_twopulses} for $M = 0.1$. While the amplitudes of both pulses are set to be equal, the asymmetric geometry of the throat requires us to precisely tune the timing of outgoing pulse ($u_{\mathrm{i}}, u_{\mathrm{f}}$) relative to the ingoing pulse. Though the left and right panels of Fig.~\ref{fig:M_twopulses} confirm instability in all cases, modest enhancements in stability are achievable: the formation of black holes and decoupling of two asymptotics regions can be delayed, analogous to some earlier results. In particular, the middle panel demonstrates that while the wormhole remains eventually unstable, strategic timing alignment of pulses temporarily sustains traversability by postponing the throat’s collapse. This implies that even for $M \neq 0$, hybrid pulse interactions can reduce but do not cure the instability, offering only limited extensions to the wormhole’s lifetime.

\begin{figure}[h]
\centering
 \includegraphics[scale=0.2]{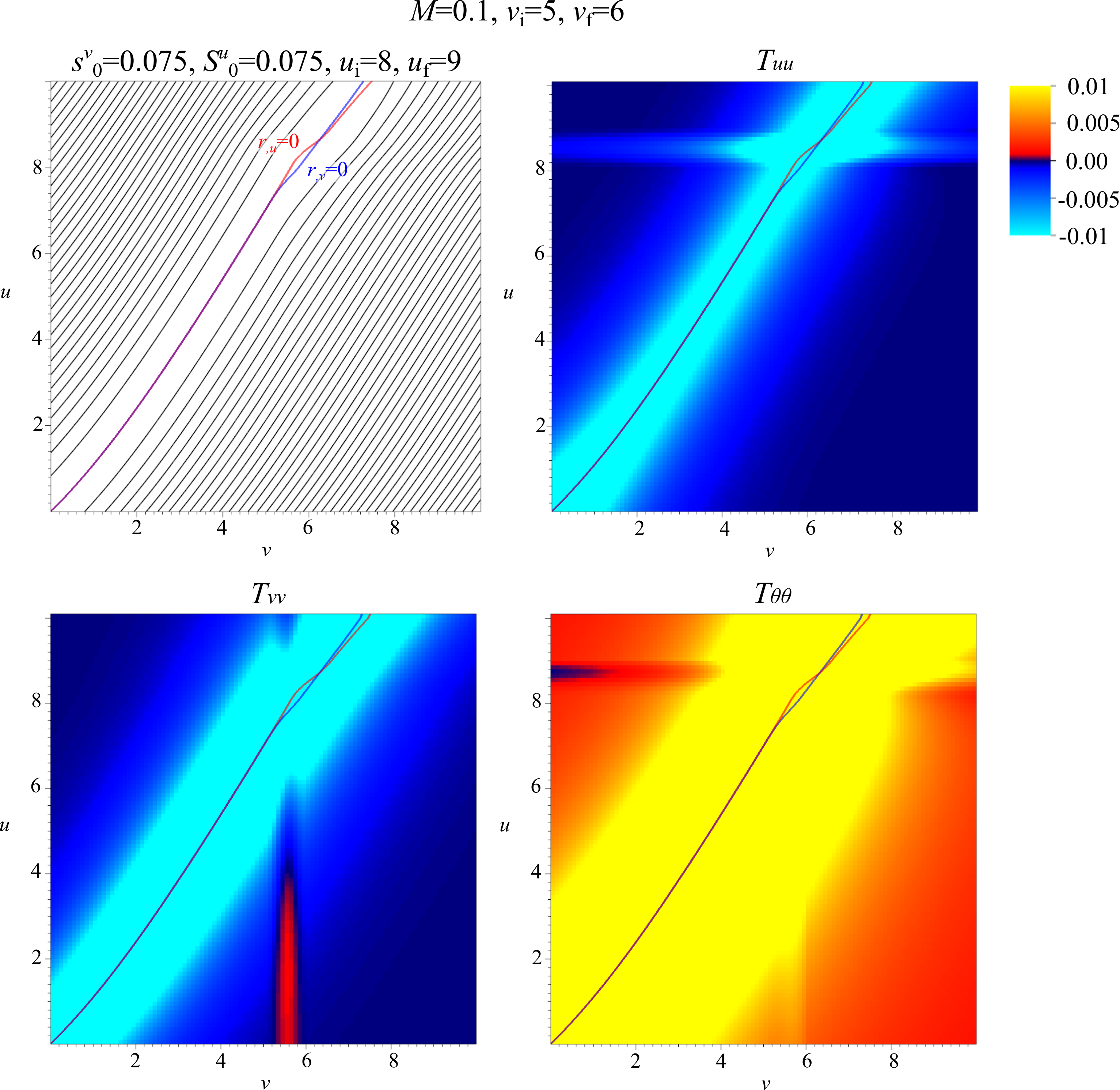}
\caption{
The components of energy-momentum tensor in clockwise direction: $T_{uu}$, $T_{\theta\theta}$, and $T_{vv}$ to describe collisions (top-left panel) of a normal scalar field with $s^v_0 = 0.075$ (the red spike in $T_{vv}$) and phantom field with $S^u_0 = 0.075$ (the blue spike in $T_{\theta\theta}$) at the throat of the massive BE wormhole with $M = 0.1$. Pulses are emitted along ingoing ($v_i=5$ to $v_f=6$) and outgoing ($u_i=7$ to $u_f=8$) null directions.
}
\label{fig:Twopulse_EMtensor}
\end{figure}

Fig.~\ref{fig:Twopulse_EMtensor} shows the components of energy-momentum tensor: $T_{uu}$, $T_{\theta\theta}$, and $T_{vv}$ to depict the dynamics of areal radius $r(u,v)$ under the collisions of two pulses from the bottom-middle panel of Fig.~\ref{fig:M_twopulses}. The appearance of a red spike in the entirely negative of $T_{vv}$ corresponds to the ingoing normal scalar field, which destroys the coincidence of the apparent horizons $r_{,v} = 0$ and $r_{,u} = 0$. Conversely, the appearance of a blue spike in entirely positive of $T_{\theta\theta}$ corresponds to the outgoing phantom field that temporarily binds the curves $r_{,v} = 0$ and $r_{,u} = 0$ before their final separation, thus exhibiting quasi-oscillatory behavior of throat. Additionally, the phantom field amplifies the violation of weak energy condition in $T_{uu}$ by expanding its negative region along the emission of the phantom field.

\section{\label{sec:con}Conclusion}

Our paper employs the double-null formalism to investigate the stability of the Bronnikov-Ellis (BE) wormhole, utilizing boundary conditions derived from the analytical solution of the Bronnikov-Ellis (BE) wormhole. The throat of the BE wormhole can be identified when the two apparent horizons $r_{,u}=0$ and $r_{,v}=0$ coincide. We probe the destabilization mechanisms of the BE wormhole by emitting ingoing and outgoing pulses composed of normal scalar and phantom fields into the wormhole spacetime. Our results demonstrate that a single propagating pulse triggers the BE wormhole to become unstable with the separation of $r_{,v}=0$ and $r_{,u}=0$ for two outcomes: BE wormhole collapses into a black hole for normal scalar field or induces inflationary expansion to decouple two asymptotic regions for the phantom field. These two scenarios can happen relatively earlier if the amplitude of pulses increases.

Next, we consider collisions between two pulses with equal amplitude emitted from ingoing and outgoing null directions; we find that the two scenarios also occur for identical pulses. However, the increase in the wormhole's mass can delay the two scenarios, temporarily prolonging the stability of the wormhole before its inevitable collapse. Notably, the strategic emission of the outgoing phantom pulse for massive BE wormhole induces the apparent horizons $r_{,u}=0$ and $r_{,v}=0$, initially separated by the ingoing normal scalar field, can temporarily coincide by exhibiting oscillating behavior. This oscillation extends the traversability of the wormhole, providing insight into reducing instability through controlled interactions of fields.

Our work addresses a longstanding issue of traversable wormholes, i.e., their instability, by proposing a hypothetical stabilization protocol: one observer traverses the wormhole while another strategically emits a phantom pulse to counteract destabilizing normal scalar field pulse. Though speculative, this highlights the potential for engineered pulse sequences to stabilize wormhole geometries, necessitating rigorous future exploration.  

Finally, the double-null formalism proves to be a robust framework for probing the stability of BE wormhole. This formalism offers more profound insights into spacetime stability than linear perturbation theory by directly addressing the nonlinear dynamics in spacetime. For future work, it enables us to study the stability of other compact objects with either analytical or numerical solutions.

\newpage

\section*{Appendix A: convergence and consistency of numerical computation}

\begin{figure}[h]
\centering
 \includegraphics[scale=0.3]{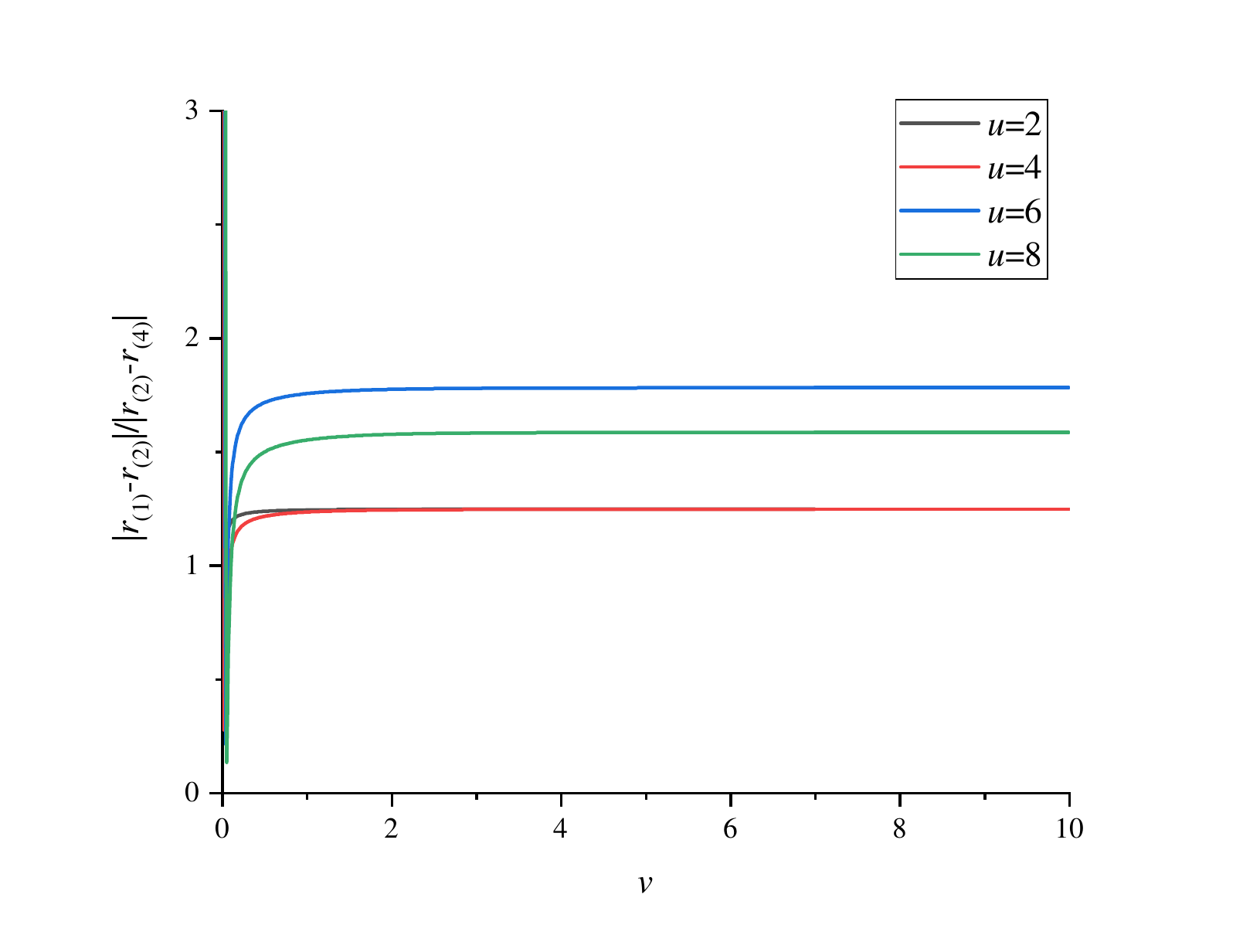}
\caption{$\left| r_{(1)} - r_{(2)} \right|/\left| r_{(2)} - r_{(4)} \right|$ for the $M = 0.1$ case.}
\label{fig:conv}
 \includegraphics[scale=0.3]{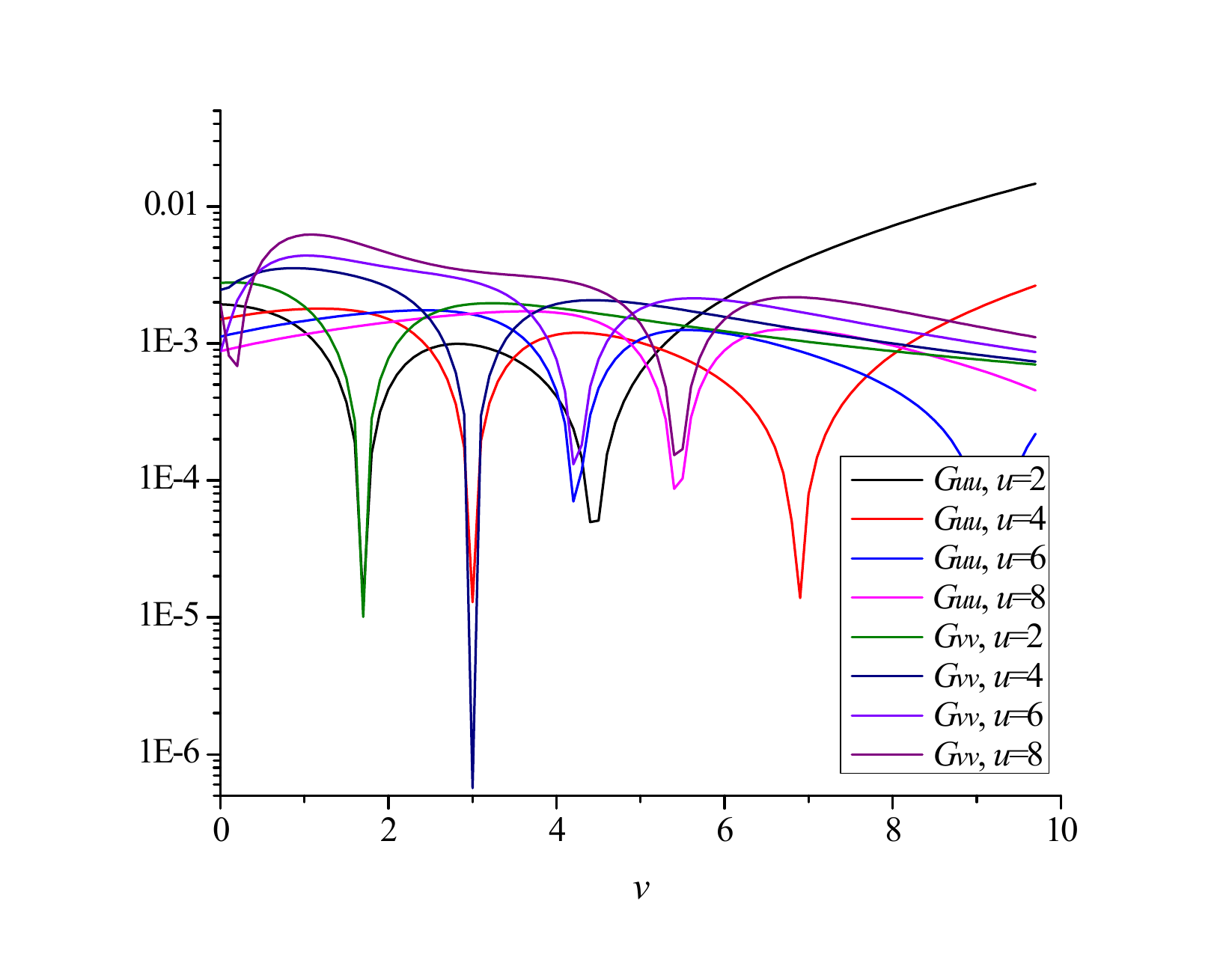}
\caption{Eq.~(\ref{eq:cons1}) (first 4 curves, denoted by $G_{uu}$) and Eq.~(\ref{eq:cons2}) (next 4 curves, denoted by $G_{vv}$) for the $M = 0.1$ case.}
\label{fig:cons}
\end{figure}

We check our numerical convergence and consistency for the case $M=0.1$, employing progressively refined resolutions: the primary resolution $r_{(1)}$, used as our default grid, is compared against a $2 \times 2$-finer simulation $r_{(2)}$ with halved step sizes along $u$ and $v$-directions, and a $4 \times 4$-finer simulation $r_{(4)}$ with step sizes reduced to one-fourth of $r_{(1)}$. By computing 
\begin{eqnarray}
\frac{\left| r_{(1)} - r_{(2)} \right|}{\left| r_{(2)} - r_{(4)} \right|}
\end{eqnarray}
for several $u = \mathrm{const.}$ slices as shown in Fig.~\ref{fig:conv}. As the results approach approximately 2, we can confirm that this simulation converges the second-order; this indicates that our numerical algorithm used the second-order method. 

To examine the consistency, we compute the following two expressions formed by the components of the Einstein equation:
\begin{eqnarray}
&& \frac{\left| G_{uu} - 8\pi T_{uu}\right|}{\left|\frac{2}{r}\right| \left( \left| f_{,u} \right| + \left| 2 f h \right| \right) + \left| 8\pi T_{uu}\right|}, \label{eq:cons1}\\
&& \frac{\left| G_{vv} - 8\pi T_{vv}\right|}{\left|\frac{2}{r}\right| \left( \left| g_{,v} \right| + \left| 2 gd \right| \right) + \left| 8\pi T_{vv}\right|}. \label{eq:cons2}
\end{eqnarray}
Notice that we did not solve the above expressions in the numerical calculation; we only use them to impose the boundary condition; hence, in principle, these constraints must vanish automatically when we have obtained the numerical solutions. However, as shown in Fig.~\ref{fig:cons} for several slices of $u = \mathrm{const.}$, their numerical residuals are orders of magnitude below unity across the entire computational domain, confirming the consistency of our numerical results.

\newpage

\section*{Acknowledgment}

DY was supported by the National Research Foundation of Korea (Grant No. : 2021R1C1C1008622, 2021R1A4A5031460). XYC was supported by the starting grant of Jiangsu University of Science and Technology (JUST). AX was supported by JUST for the research trip to Pusan National University. We also acknowledged having valuable discussions with Jakob Hansen and Igor Novikov during the early stage of this research.


\begin{thebibliography}{200}


\bibitem{Caldwell:1999ew} 
  R.~R.~Caldwell,
  Phys.\ Lett.\ B {\bf 545}, 23 (2002)
  [astro-ph/9908168].

\bibitem{Carroll:2003st} 
  S.~M.~Carroll, M.~Hoffman and M.~Trodden,
  Phys.\ Rev.\ D {\bf 68}, 023509 (2003)
  [astro-ph/0301273].

\bibitem{Gibbons:2003yj} 
  G.~W.~Gibbons,
  hep-th/0302199.


\bibitem{Hannestad:2005fg} 
  S.~Hannestad,
  Int.\ J.\ Mod.\ Phys.\ A {\bf 21}, 1938 (2006)
  [astro-ph/0509320].


\bibitem{Ellis:1973yv}
  H.~G.~Ellis,
  J.\ Math.\ Phys.\  {\bf 14}, 104-118 (1973).

\bibitem{Ellis:1979bh}
  H.~G.~Ellis,
  Gen.\ Rel.\ Grav.\  {\bf 10}, 105-123 (1979).

\bibitem{Bronnikov:1973fh}
  K.~A.~Bronnikov,
  Acta Phys.\ Polon.\  {\bf B4}, 251-266 (1973).

\bibitem{Torii:2013xba} 
  T.~Torii and H.~a.~Shinkai,
  Phys.\ Rev.\ D {\bf 88}, 064027 (2013)
  [arXiv:1309.2058 [gr-qc]].

\bibitem{Kashargin:2007mm}
  P.~E.~Kashargin and S.~V.~Sushkov,
  Grav.\ Cosmol.\  {\bf 14}, 80 (2008).
  [arXiv:0710.5656 [gr-qc]].

\bibitem{Kashargin:2008pk}
  P.~E.~Kashargin and S.~V.~Sushkov,
  Phys.\ Rev.\ D {\bf 78}, 064071 (2008).
  [arXiv:0809.1923 [gr-qc]].

\bibitem{Kleihaus:2014dla}
  B.~Kleihaus and J.~Kunz,
  Phys.\ Rev.\ D {\bf 90} (2014) 121503
  [arXiv:1409.1503 [gr-qc]].

\bibitem{Chew:2016epf}
  X.~Y.~Chew, B.~Kleihaus and J.~Kunz,
  Phys.\ Rev.\ D {\bf 94} (2016) no.10,  104031
  [arXiv:1608.05253 [gr-qc]].

\bibitem{Chew:2018vjp}
X.~Y.~Chew, B.~Kleihaus and J.~Kunz,
Phys. Rev. D \textbf{97} (2018) no.6, 064026
[arXiv:1802.00365 [gr-qc]].


\bibitem{Dzhunushaliev:2013jja} 
  V.~Dzhunushaliev, V.~Folomeev, B.~Kleihaus, J.~Kunz and E.~Radu,
  Phys.\ Rev.\ D {\bf 88}, 124028 (2013).
  [arXiv:1309.2448 [gr-qc]].

  
\bibitem{Dzhunushaliev:2014bya} 
  V.~Dzhunushaliev, V.~Folomeev, C.~Hoffmann, B.~Kleihaus and J.~Kunz,
  Phys.\ Rev.\ D {\bf 90}, no. 12, 124038 (2014)
 [arXiv:1409.6978 [gr-qc]].

\bibitem{Hoffmann:2018oml} 
  C.~Hoffmann, T.~Ioannidou, S.~Kahlen, B.~Kleihaus and J.~Kunz,
  Phys.\ Rev.\ D {\bf 97}, no. 12, 124019 (2018)
 [arXiv:1803.11044 [gr-qc]].

\bibitem{Ding:2023syj}
P.~B.~Ding, T.~X.~Ma, T.~F.~Fang and Y.~Q.~Wang,
JHEP \textbf{04} (2024), 033
[arXiv:2305.19819 [gr-qc]].



\bibitem{Dzhunushaliev:2011xx} 
  V.~Dzhunushaliev, V.~Folomeev, B.~Kleihaus and J.~Kunz,
  JCAP {\bf 1104}, 031 (2011)
 [arXiv:1102.4454 [astro-ph.GA]].
 
 
\bibitem{Aringazin:2014rva} 
  A.~Aringazin, V.~Dzhunushaliev, V.~Folomeev, B.~Kleihaus and J.~Kunz,
  JCAP {\bf 1504}, no. 04, 005 (2015)
 [arXiv:1412.3194 [gr-qc]].

\bibitem{Dzhunushaliev:2022elv}
V.~Dzhunushaliev, V.~Folomeev, B.~Kleihaus and J.~Kunz,
Phys. Rev. D \textbf{107} (2023) no.4, 044060
[arXiv:2210.04425 [gr-qc]].


\bibitem{Chew:2019lsa}
X.~Y.~Chew, V.~Dzhunushaliev, V.~Folomeev, B.~Kleihaus and J.~Kunz,
Phys. Rev. D \textbf{100} (2019) no.4, 044019
[arXiv:1906.08742 [gr-qc]].

\bibitem{Chew:2020svi}
X.~Y.~Chew and K.~G.~Lim,
Phys. Rev. D \textbf{102} (2020) no.12, 124068
[arXiv:2009.13334 [gr-qc]].

\bibitem{Chew:2021vxh}
X.~Y.~Chew and K.~G.~Lim,
Phys. Rev. D \textbf{105} (2022) no.8, 084058
[arXiv:2109.00262 [gr-qc]].

\bibitem{Martinez:2020hjm}
C.~Martinez and M.~Nozawa,
Phys. Rev. D \textbf{103} (2021) no.2, 024003
[arXiv:2010.05183 [gr-qc]].

\bibitem{DuttaRoy:2019hij}
P.~Dutta Roy, S.~Aneesh and S.~Kar,
Eur. Phys. J. C \textbf{80} (2020) no.9, 850
[arXiv:1910.08746 [gr-qc]].


\bibitem{Sharma:2022dbx}
V.~Sharma and S.~Ghosh,
Eur. Phys. J. Plus \textbf{137} (2022) no.8, 881
[arXiv:2205.08865 [gr-qc]].

\bibitem{Huang:2020qmn}
H.~Huang, H.~L\"u and J.~Yang,
Class. Quant. Grav. \textbf{39} (2022) no.18, 185009
[arXiv:2010.00197 [gr-qc]].

\bibitem{Blazquez-Salcedo:2020nsa}
J.~L.~Bl\'azquez-Salcedo, X.~Y.~Chew, J.~Kunz and D.~H.~Yeom,
Eur. Phys. J. C \textbf{81} (2021) no.9, 858
[arXiv:2012.06213 [gr-qc]].


\bibitem{Karakasis:2021tqx}
T.~Karakasis, E.~Papantonopoulos and C.~Vlachos,
Phys. Rev. D \textbf{105} (2022) no.2, 024006
[arXiv:2107.09713 [gr-qc]].



\bibitem{Filho:2023yly}
A.~A.~A.~Filho, J.~Furtado, J.~A.~A.~S.~Reis and J.~E.~G.~Silva,
Class. Quant. Grav. \textbf{40} (2023) no.24, 245001
[arXiv:2302.05492 [gr-qc]].

\bibitem{Lin:2022azn}
K.~Lin and W.~L.~Qian,
Eur. Phys. J. C \textbf{82} (2022) no.6, 529
[arXiv:2203.03081 [gr-qc]].

\bibitem{Gonzalez:2022ote}
P.~A.~Gonz\'alez, E.~Papantonopoulos, \'A.~Rinc\'on and Y.~V\'asquez,
Phys. Rev. D \textbf{106} (2022) no.2, 024050
[arXiv:2205.06079 [gr-qc]].

\bibitem{Hao:2024hba}
C.~H.~Hao, X.~Su and Y.~Q.~Wang,
[arXiv:2404.11002 [gr-qc]].

\bibitem{Gao:2024lrb}
C.~Gao and J.~Qiu,
Universe \textbf{10} (2024) no.8, 328
[arXiv:2407.14184 [gr-qc]].

\bibitem{Yang:2024prm}
H.~Yang, Z.~W.~Xia and Y.~G.~Miao,
[arXiv:2406.00377 [gr-qc]].

\bibitem{Ding:2024qrf}
C.~Ding, C.~Liu, Y.~Xiao and J.~Chen,
[arXiv:2407.16916 [gr-qc]].

\bibitem{Liu:2025foo}
H.~C.~Liu and R.~X.~Miao,
[arXiv:2501.02324 [hep-th]].

\bibitem{Bouhmadi-Lopez:2021zwt}
M.~Bouhmadi-L\'opez, C.~Y.~Chen, X.~Y.~Chew, Y.~C.~Ong and D.~h.~Yeom,
JCAP \textbf{10} (2021), 059
[arXiv:2108.07302 [gr-qc]].

\bibitem{Barros:2018lca}
B.~J.~Barros and F.~S.~N.~Lobo,
Phys. Rev. D \textbf{98} (2018) no.4, 044012
[arXiv:1806.10488 [gr-qc]].

\bibitem{Tangphati:2023uxt}
T.~Tangphati, B.~Chaihao, D.~Samart, P.~Channuie and D.~Momeni,
Nucl. Phys. B \textbf{999} (2024), 116446
[arXiv:2307.13968 [gr-qc]].

\bibitem{Blazquez-Salcedo:2019uqq}
J.~L.~Bl\'azquez-Salcedo and C.~Knoll,
Eur. Phys. J. C \textbf{80} (2020) no.2, 174
[arXiv:1910.03565 [gr-qc]].

\bibitem{Churilova:2021tgn}
M.~S.~Churilova, R.~A.~Konoplya, Z.~Stuchlik and A.~Zhidenko,
JCAP \textbf{10} (2021), 010
[arXiv:2107.05977 [gr-qc]].

\bibitem{Wang:2022aze}
Y.~Q.~Wang, S.~W.~Wei and Y.~X.~Liu,
[arXiv:2206.12250 [gr-qc]].



\bibitem{Kanti:2011yv}
P.~Kanti, B.~Kleihaus and J.~Kunz,
Phys. Rev. D \textbf{85} (2012), 044007
[arXiv:1111.4049 [hep-th]].

\bibitem{Ibadov:2020btp}
R.~Ibadov, B.~Kleihaus, J.~Kunz and S.~Murodov,
Phys. Rev. D \textbf{102} (2020) no.6, 064010
[arXiv:2006.13008 [gr-qc]].

\bibitem{Canate:2023mge}
P.~Ca\~nate,
Phys. Rev. D \textbf{108} (2023) no.10, 104048
[arXiv:2310.08758 [gr-qc]].

\cite{Ilyas:2023rde}
\bibitem{Ilyas:2023rde}
M.~Ilyas and K.~Bamba,
JCAP \textbf{10} (2023), 038
[arXiv:2308.11004 [gr-qc]].

\bibitem{Gonzalez:2008wd}
  J.~A.~Gonzalez, F.~S.~Guzman, and O.~Sarbach,
  Class.\ Quant.\ Grav.\  {\bf 26}, 015010 (2009).
  [arXiv:0806.0608 [gr-qc]].

\bibitem{Gonzalez:2008xk}
  J.~A.~Gonzalez, F.~S.~Guzman, and O.~Sarbach,
  Class.\ Quant.\ Grav.\  {\bf 26}, 015011 (2009).
  [arXiv:0806.1370 [gr-qc]].

\bibitem{Bronnikov:2012ch}
K.~A.~Bronnikov, R.~A.~Konoplya and A.~Zhidenko,
Phys. Rev. D \textbf{86} (2012), 024028
[arXiv:1205.2224 [gr-qc]].

\bibitem{Blazquez-Salcedo:2018ipc} 
  J.~L.~Blázquez-Salcedo, X.~Y.~Chew and J.~Kunz,
  Phys.\ Rev.\ D {\bf 98}, no. 4, 044035 (2018)
  [arXiv:1806.03282 [gr-qc]].

\bibitem{Azad:2023iju}
B.~Azad, J.~L.~Blazquez-Salcedo, F.~S.~Khoo and J.~Kunz,
Phys. Lett. B \textbf{848} (2024), 138349
[arXiv:2301.05243 [gr-qc]].

\bibitem{Azad:2024axu}
B.~Azad, J.~L.~Bl\'azquez-Salcedo, F.~S.~Khoo and J.~Kunz,
Phys. Rev. D \textbf{109} (2024) no.12, 124051
[arXiv:2403.08387 [gr-qc]].




\bibitem{Kim:2008zzj}
S.~W.~Kim,
Prog. Theor. Phys. Suppl. \textbf{172} (2008), 21-29


\bibitem{Khoo:2024yeh}
F.~S.~Khoo, B.~Azad, J.~L.~Bl\'azquez-Salcedo, L.~M.~Gonz\'alez-Romero, B.~Kleihaus, J.~Kunz and F.~Navarro-L\'erida,
Phys. Rev. D \textbf{109} (2024) no.8, 084013
[arXiv:2401.02898 [gr-qc]].


\bibitem{Azad:2022qqn}
B.~Azad, J.~L.~Bl\'azquez-Salcedo, X.~Y.~Chew, J.~Kunz and D.~h.~Yeom,
Phys. Rev. D \textbf{107} (2023) no.8, 084024
[arXiv:2212.12601 [gr-qc]].

\bibitem{Batic:2025hgp}
D.~Batic and D.~Dutykh,
Eur. Phys. J. C \textbf{85} (2025) no.2, 144
[arXiv:2502.05486 [gr-qc]].






\bibitem{Nakonieczna:2018tih}
A.~Nakonieczna, \L{}.~Nakonieczny and D.~Yeom,
Int. J. Mod. Phys. D \textbf{28} (2018) no.03, 1930006
[arXiv:1805.12362 [gr-qc]].

\bibitem{Waugh:1986jh}
B.~Waugh and K.~Lake,
Phys. Rev. D \textbf{34}, 2978-2984 (1986).

\bibitem{Hong:2008mw}
S.~E.~Hong, D.~Hwang, E.~D.~Stewart and D.~Yeom,
Class. Quant. Grav. \textbf{27} (2010), 045014
[arXiv:0808.1709 [gr-qc]].

\bibitem{Hwang:2012nn}
D.~Hwang, B.~H.~Lee and D.~Yeom,
JCAP \textbf{01} (2013), 005
[arXiv:1210.6733 [gr-qc]].

\bibitem{Chew:2023upu}
X.~Y.~Chew and D.~h.~Yeom,
J. Korean Phys. Soc. \textbf{85} (2024) no.12, 1050-1061
[arXiv:2308.09225 [gr-qc]].

\bibitem{Hansen:2009kn}
J.~Hansen, D.~Hwang and D.~Yeom,
JHEP \textbf{11} (2009), 016
[arXiv:0908.0283 [gr-qc]].

\bibitem{Hwang:2012pj}
D.~Hwang, B.~H.~Lee, W.~Lee and D.~Yeom,
JCAP \textbf{07} (2012), 003
[arXiv:1201.6109 [gr-qc]].

\bibitem{Doroshkevich:2008xm}
A.~Doroshkevich, J.~Hansen, I.~Novikov and A.~Shatskiy,
Int. J. Mod. Phys. D \textbf{18} (2009), 1665-1691
[arXiv:0812.0702 [gr-qc]].

\bibitem{Hwang:2011mn}
D.~Hwang, H.~B.~Kim and D.~Yeom,
Class. Quant. Grav. \textbf{29} (2012), 055003
[arXiv:1105.1371 [gr-qc]].

\bibitem{Ashtekar:2004cn}
A.~Ashtekar and B.~Krishnan,
Living Rev. Rel. \textbf{7}, 10 (2004)
[arXiv:gr-qc/0407042 [gr-qc]].

\end{thebibliography}
\end{document}